\documentclass[12pt,doublespacing]{article}

\usepackage{fullpage}
\usepackage{setspace}
\usepackage{authblk}

\usepackage{cite}
\usepackage{hyperref} 

\usepackage{graphicx}

\usepackage{amsmath}
\usepackage{amssymb}
\allowdisplaybreaks

\usepackage{float}
\usepackage{subfig}
\usepackage{makecell}
\usepackage{rotating}

\usepackage{enumitem}

\begin{document}

\title{A Single-Trace Surface Integral Equation Solver for Simulation of Open Bianisotropic Metasurfaces Described by Generalized Sheet Transition Conditions}

\author[1]{Sebastian Celis Sierra}
\author[1]{Junze Shao}
\author[2]{Ran Zhao}
\author[3]{Rui Chen}
\author[1]{Partha Mondal}
\author[1]{Hakan Bagci\vspace{0.5cm}}

\affil[1]{S. Celis Sierra, J. Shao, P. Mondal, and H. Bagci are with the Electrical and Computer Engineering Program, Computer, Electrical, and Mathematical Science and Engineering Division, King Abdullah University of Science and Technology (KAUST), Thuwal 23955-6900, Saudi Arabia (e-mails: \{sebastian.celissierra, junze.shao, partha.mondal, hakan.bagci\}@kaust.edu.sa).\vspace{0.5cm}}

\affil[2]{R. Zhao was with the Electrical and Computer Engineering (ECE) Program, Computer, Electrical, and Mathematical Science and Engineering (CEMSE) Division, King Abdullah University of Science and Technology (KAUST), Thuwal 23955-6900, Saudi Arabia. He is now with the School of Electronic Science and Engineering, University of Electronic Science and Technology of China (UESTC), Chengdu 61173, China (e-mail: ran.zhao@uestc.edu.cn).\vspace{0.5cm}}

\affil[3]{R. Chen was with the Electrical and Computer Engineering (ECE) Program, Computer, Electrical, and Mathematical Science and Engineering (CEMSE) Division, King Abdullah University of Science and Technology (KAUST), Thuwal 23955-6900, Saudi Arabia. He is now with the School of Microelectronics (School of Integrated Circuits), Nanjing University of Science and Technology, Nanjing 210094, China (e-mail: rui.chen@njust.edu.cn).}

\date{}
\maketitle
\newpage

\begin{abstract}
A single-trace surface integral equation (SIE) solver incorporating generalized sheet transition conditions (GSTCs) is presented for the simulation of three-dimensional (3D) open bianisotropic metasurfaces. The metasurface is modeled as an infinitesimally thin, non-enclosing sheet across which the GSTCs enforce the electromagnetic field discontinuities through four surface susceptibility tensors. The proposed solver uses a single set of equivalent surface currents on the sheet, in place of the two sets used by prior multi-trace formulations. The scattered fields on both faces of the sheet, expressed through SIE operators acting on these currents, are substituted into the GSTCs. The resulting system of equations is then discretized using Rao--Wilton--Glisson basis functions. This solver models an open metasurface directly, without an artificial closure, and applies to both planar and curved geometries. It is validated against analytical solutions for polarization rotation and perfect reflection, and is used to model a realistic broadband absorber whose susceptibility tensors are retrieved from full-wave simulation data. A direct comparison shows that the single-trace formulation attains lower error than a multi-trace formulation while using significantly fewer unknowns.
\end{abstract}

\section{Introduction}\label{sec:Intro}
A metasurface is an electrically thin structure formed by an array of subwavelength unit cells. Each cell is designed to impose a local scattering response, and the spatial variation of these responses across the array synthesizes a prescribed transformation of the incident electromagnetic field, manipulating its phase, amplitude, polarization, and propagation direction~\cite{MS_Overview,Caloz_Book}. Such control of electromagnetic fields enables applications ranging from beamforming~\cite{beamforming2022} and beam steering~\cite{beamsteering2020, beamsteering2023} in wireless communication channels to radar cross section reduction~\cite{metasurface_rcs_reduction2020, metasurface_rcs_reduction2022, metasurface_rcs_reduction2023} in radar systems, many of which call for metasurfaces deployed on electrically large platforms. 

Predicting the performance of such a deployed metasurface is challenging. Traditional simulation methods often rely on full-wave analysis of a single unit cell under the assumption of infinite periodicity~\cite{Caloz_Book}. These methods are well suited to design, where the unit-cell geometry is tuned to realize a prescribed local response. However, they are not well suited to modeling the behavior of the resulting device in its operating environment: a deployed metasurface has a finite aperture, may conform to a curved platform, and is illuminated by a beam rather than a plane wave, none of which can be accounted for by a periodic model. Capturing these effects requires simulating the metasurface together with its platform, for which a direct full-wave discretization of the physical structure is computationally expensive, since the subwavelength features of the unit cells must be resolved across the metasurface, whose aperture spans an electrically large domain.

This cost is avoided by replacing the physical metasurface with an equivalent infinitesimally thin sheet across which generalized sheet transition conditions (GSTCs) are enforced~\cite{Kuester, idemen2011discontinuities}. The GSTCs relate the fields on both sides of the sheet through four surface susceptibility tensors $\bar{\bar{\chi}}_{\mathrm{ab}}$, $\mathrm{a,b}\in\{\mathrm{e,m}\}$~\cite{Susceptibility_Tensors, Synthesis_Spherical, achouri2018, liu2019generalized}, which represent the full bianisotropic response of the metasurface. This approach eliminates the need for fine-scale discretization of the subwavelength unit cells.

For this reason, the GSTCs have been incorporated into a range of full-wave solvers. Differential equation-based formulations include finite-difference frequency-domain (FDFD)~\cite{FDFD}, finite element method (FEM)~\cite{FEM_Caloz}, finite-difference time-domain (FDTD)~\cite{FDTD_Jia_Xiao, FDTD_Caloz, FDTD_Wu}, and discontinuous Galerkin time-domain (DGTD)~\cite{DGTD_GSTC1,DGTD_GSTC2,chen_2023}. These implementations, however, inherit the intrinsic limitations of differential-equation schemes: the entire computational domain must be discretized, its truncation requires absorbing boundary conditions or perfectly matched layers, and numerical phase dispersion accumulates over electrically large domains~\cite{jin2015theory, peterson1998}. These restrictions become especially severe when the metasurface is embedded in an extended background medium.

In contrast, surface integral equation (SIE) solvers discretize only the surfaces, which significantly reduces the number of unknowns relative to volumetric discretization. Since the fields are represented using the background medium Green function, the radiation condition is implicitly satisfied, and the absence of a volumetric grid eliminates the accumulation of numerical phase dispersion~\cite{jin2015theory, peterson1998}. However, to date, most SIE solvers that incorporate GSTCs for modeling metasurfaces have been limited to two-dimensional (2D) problems~\cite{Gupta_Coordinates, MoM_Scott, MoM_Sandeep, celis_2023a, Gupta_1, MarioP}. 

\begin{figure}[!ht]
\centering
\includegraphics[width=0.9\columnwidth]{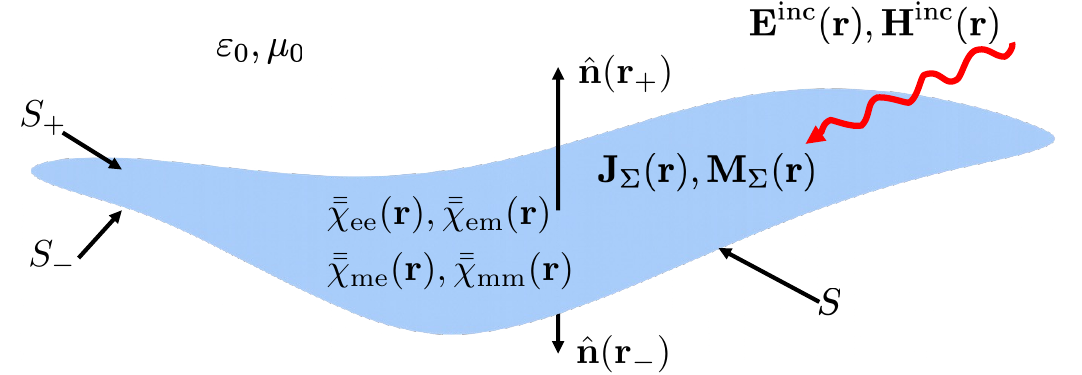}
\caption{Description of the electromagnetic scattering problem, where $S$ represents the open metasurface.}
\label{fig:Open_metasurface3D}
\end{figure}

Extensions to three-dimensional (3D) configurations have only recently emerged, and remain limited in scope~\cite{journal1_Celis_2026,celis_2023b, budhu_2024, Kim2025}. The formulations in~\cite{journal1_Celis_2026, celis_2023b, budhu_2024} are multi-trace: two independent sets of equivalent currents are defined on the interior and exterior sides of a topologically closed surface and coupled through the GSTCs. In these formulations, an open metasurface is modeled by extending it into a topologically closed surface and enforcing the nonzero susceptibility tensors only on the portion representing the metasurface, at the cost of a substantially increased number of unknowns. The SIE-GSTC solver in~\cite{Kim2025} instead treats a perfectly absorbing metasurface mounted on a closed perfect electrically conducting (PEC) body, where the fields vanish in the interior region. This assumption precludes a sheet supporting independent fields on both sides. Furthermore, the metasurfaces considered in~\cite{journal1_Celis_2026, celis_2023b, Kim2025} are monoanisotropic, so the magnetoelectric coupling tensors $\bar{\bar{\chi}}_{\mathrm{em}}$ and $\bar{\bar{\chi}}_{\mathrm{me}}$ are absent. Here, an open metasurface refers to a non-enclosing, infinitesimally thin sheet that does not bound a volumetric region and across which the GSTCs enforce the field discontinuities (see Fig.~\ref{fig:Open_metasurface3D}). Its two faces are therefore exposed to the same background medium, and the equivalent currents on them cannot be assigned to separate regions.

In this work, these limitations are addressed by introducing a single-trace SIE-GSTC solver for 3D open metasurfaces. The metasurface is modeled as an infinitesimally thin, non-enclosing sheet whose response is represented by four bianisotropic susceptibility tensors, synthesized using established retrieval procedures~\cite{journal1_Celis_2026,Susceptibility_Tensors,Synthesis_Spherical,achouri2018, liu2019generalized}. The single-trace formulation defines one set of equivalent currents on the sheet, in place of the two sets used by prior multi-trace formulations~\cite{journal1_Celis_2026, celis_2023b, budhu_2024}. This allows for direct modeling of an open metasurface, without the need for an artificial closure. The scattered fields on both faces of the sheet are expressed using SIE operators acting on this single set of currents. Their average and difference are substituted into the GSTCs, yielding the single-trace SIE-GSTC system, which is discretized using Rao--Wilton--Glisson (RWG) basis functions defined on the triangular mesh of the sheet~\cite{RWG}. The formulation accommodates curved as well as planar metasurfaces. Because the susceptibility tensors generally vary across the surface, their surface divergence introduces an additional term, which is treated consistently through its line-integral contributions along the edges of the mesh triangles.

A preliminary version of this solver is reported in~\cite{celis_2024c}. The solver is validated on monoanisotropic and bianisotropic transformations, including polarization rotation and perfect reflection, on both planar and curved geometries. Numerical results further show that it reproduces the response of a realistic broadband plasmonic absorber whose susceptibility tensors are retrieved from full-wave FDTD data, and that it attains lower error than a multi-trace formulation while using significantly fewer unknowns. 

The remainder of this paper is structured as follows: Section~\ref{sec:Formulation} presents the single-trace SIE-GSTC formulation and its discretization. Section~\ref{sec:Numerical_Examples} reports numerical results validating the solver, and Section~\ref{sec:Conclusions} concludes the paper. 
\section{Formulation}\label{sec:Formulation}

\subsection{SIE-GSTC System}\label{sec:SIE-GSTCs}
Let $S$ denote an infinitesimally thin surface embedded in an unbounded homogeneous background medium with permittivity $\varepsilon_0$, permeability $\mu_0$, wave impedance $\eta_0=\sqrt{\mu_0/\varepsilon_0}$, wavenumber $k_0$, and wavelength $\lambda_0$ (see Fig.~\ref{fig:Open_metasurface3D}). The surface $S$ is illuminated by an incident electromagnetic field, where $\mathbf{E}^{\mathrm{inc}}(\mathbf{r})$ and $\mathbf{H}^{\mathrm{inc}}(\mathbf{r})$ denote the electric and magnetic fields, respectively. The angular frequency of excitation is denoted by $\omega$.  

Since $S$ is an open, infinitesimally thin sheet that does not bound a volumetric region, the classical interior–exterior decomposition typical of SIE formulations for penetrable scatterers is not applicable~\cite{journal1_Celis_2026, jin2015theory}. The surface $S$ has two faces, $S_{+}$ and $S_{-}$, where $\mathbf{r}_{\pm}$ denotes point $\mathbf{r}$ on $S$ approached from face $S_{\pm}$. The unit normals of these faces satisfy $\hat{\mathbf{n}}(\mathbf{r}_{+}) = -\hat{\mathbf{n}}(\mathbf{r}_{-})$. In the single-trace formulation proposed here, a single set of equivalent electric and magnetic surface currents $\mathbf{J}_{\Sigma}(\mathbf{r})$ and $\mathbf{M}_{\Sigma}(\mathbf{r})$ is introduced as the unknowns for $\mathbf{r} \in S$:
\begin{equation}
\label{eq:sum_current_J}
    \mathbf{J}_{\Sigma}(\mathbf{r}) = \mathbf{J}(\mathbf{r}_{+}) + \mathbf{J}(\mathbf{r}_{-})
\end{equation}
\begin{equation}
\label{eq:sum_current_M}
\mathbf{M}_{\Sigma}(\mathbf{r}) = \mathbf{M}(\mathbf{r}_{+}) + \mathbf{M}(\mathbf{r}_{-})
\end{equation}
Here, $\mathbf{J}(\mathbf{r}_{\pm}) = \hat{\mathbf{n}}(\mathbf{r}_{\pm}) \times \mathbf{H}(\mathbf{r}_{\pm})$ and $\mathbf{M}(\mathbf{r}_{\pm})=-\hat{\mathbf{n}}(\mathbf{r}_{\pm}) \times \mathbf{E}(\mathbf{r}_{\pm})$ are the equivalent electric and magnetic surface currents on face $S_{\pm}$, respectively. $\mathbf{E}(\mathbf{r}_{\pm})$ and $\mathbf{H}(\mathbf{r}_{\pm})$ denote the total electric and magnetic fields on face $S_{\pm}$, respectively. The reference unit normal is defined as $\hat{\mathbf{n}}(\mathbf{r}) = \hat{\mathbf{n}}(\mathbf{r}_{+})$. Then, $\mathbf{J}_{\Sigma}(\mathbf{r})$ and $\mathbf{M}_{\Sigma}(\mathbf{r})$ are related to $\mathbf{E}(\mathbf{r}_{\pm})$ and $\mathbf{H}(\mathbf{r}_{\pm})$, via the jump conditions across $S$:
\begin{equation}
\label{eq:jump_currents_J}
    \mathbf{J}_{\Sigma}(\mathbf{r}) = \hat{\mathbf{n}}(\mathbf{r}) \times \big[ \mathbf{H}(\mathbf{r}_{+}) - \mathbf{H}(\mathbf{r}_{-}) \big]
\end{equation}
\begin{equation}
\label{eq:jump_currents_M}
    \mathbf{M}_{\Sigma}(\mathbf{r}) = -\hat{\mathbf{n}}(\mathbf{r}) \times \big[ \mathbf{E}(\mathbf{r}_{+}) - \mathbf{E}(\mathbf{r}_{-}) \big]
\end{equation}
for $\mathbf{r} \in S$. These jump conditions are governed by the GSTCs enforced on $S$, which describe the electromagnetic response of the metasurface and read~\cite{Susceptibility_Tensors}:
\begin{equation}
\label{eq:GSTC_H}
    \hat{\mathbf{n}}(\mathbf{r}) \times \big[\mathbf{H}(\mathbf{r}_{+})-\mathbf{H}(\mathbf{r}_{-})\big]= -\mathrm{j}\omega \hat{\mathbf{n}}(\mathbf{r})\times \hat{\mathbf{n}}(\mathbf{r})\times \mathbf{P}^{\mathrm{e}}(\mathbf{r})-\hat{\mathbf{n}}(\mathbf{r}) \times \nabla_{\mathrm{s}}\big[\hat{\mathbf{n}}(\mathbf{r}) \cdot \mathbf{P}^\mathrm{m}(\mathbf{r})\big]
\end{equation}
\begin{equation}
\label{eq:GSTC_E}
    \hat{\mathbf{n}}(\mathbf{r}) \times \big[\mathbf{E}(\mathbf{r}_{+})-\mathbf{E}(\mathbf{r}_{-})\big]= -\mathrm{j}\omega \mu_0 \hat{\mathbf{n}}(\mathbf{r}) \times \hat{\mathbf{n}}(\mathbf{r}) \times\mathbf{P}^{\mathrm{m}}(\mathbf{r})-\frac{1}{\varepsilon_0}\hat{\mathbf{n}}(\mathbf{r}) \times\nabla_{\mathrm{s}}\big[\hat{\mathbf{n}}(\mathbf{r}) \cdot \mathbf{P}^\mathrm{e}(\mathbf{r})\big]
\end{equation}
for $\mathbf{r} \in S$. Here, $\nabla_{\mathrm{s}}f(\mathbf{r})= [\bar{\bar{I}} - \hat{\mathbf{n}}(\mathbf{r})\hat{\mathbf{n}}(\mathbf{r})]\cdot\nabla f(\mathbf{r})$ is the surface gradient operator, $\bar{\bar{I}}$ is the identity dyadic, and $\mathbf{P}^{\mathrm{e}}(\mathbf{r})$ and $\mathbf{P}^{\mathrm{m}}(\mathbf{r})$ denote the surface electric and magnetic polarization densities, respectively. The polarization densities $\mathbf{P}^{\mathrm{e}}(\mathbf{r})$ and $\mathbf{P}^{\mathrm{m}}(\mathbf{r})$ are expressed as
\begin{equation}
\label{eq:pol_e}
   \mathbf{P}^{\mathrm{e}}(\mathbf{r})=\frac{1}{2}\varepsilon_0 \bar {\bar {\chi}}_{\mathrm{ee}}(\mathbf{r})\cdot \big[\mathbf{E}(\mathbf{r}_{+})+\mathbf{E}(\mathbf{r}_{-})\big]+\frac{1}{2}\sqrt{\mu_0\varepsilon_0} \bar {\bar {\chi}}_{\mathrm{em}}(\mathbf{r})\cdot \big[\mathbf{H}(\mathbf{r}_{+})+\mathbf{H}(\mathbf{r}_{-})\big]
\end{equation}
\begin{equation}
\label{eq:pol_m}
   \mathbf{P}^{\mathrm{m}}(\mathbf{r})=\frac{1}{2}\sqrt{\frac{\varepsilon_0}{\mu_0}} \bar {\bar {\chi}}_{\mathrm{me}}(\mathbf{r})\cdot \big[\mathbf{E}(\mathbf{r}_{+})+\mathbf{E}(\mathbf{r}_{-})\big]+\frac{1}{2}\bar {\bar {\chi}}_{\mathrm{mm}}(\mathbf{r})\cdot \big[\mathbf{H}(\mathbf{r}_{+})+\mathbf{H}(\mathbf{r}_{-})\big]      
\end{equation}
for $\mathbf{r} \in S$. Here, $\bar{\bar{\chi}}_{\mathrm{ab}}(\mathbf{r})$, $\mathrm{a,b}\in\{\mathrm{e,m}\}$, are the surface susceptibility tensors. In this work, it is assumed that the susceptibility tensors are purely tangential, i.e.,
\begin{equation}
\label{eq:chi_tangential}
\bar{\bar{\chi}}_{\mathrm{ab}}(\mathbf{r})\cdot\hat{\mathbf{n}}(\mathbf{r})= \hat{\mathbf{n}}(\mathbf{r})\cdot\bar{\bar{\chi}}_{\mathrm{ab}}(\mathbf{r})= \mathbf{0}, \; \mathrm{a,b}\in\{\mathrm{e,m}\}.
\end{equation}
Consequently, $\hat{\mathbf{n}}(\mathbf{r})\cdot\mathbf{P}^{\mathrm{e}}(\mathbf{r})=0$ and $\hat{\mathbf{n}}(\mathbf{r})\cdot\mathbf{P}^{\mathrm{m}}(\mathbf{r})=0$ in~\eqref{eq:GSTC_H}--\eqref{eq:GSTC_E} and \eqref{eq:pol_e}--\eqref{eq:pol_m}.
This assumption is consistent with the wave transformations considered herein, which do not require degrees of freedom associated with normal field components~\cite{Achouri_Angular_Scattering, Gupta_Coordinates, journal1_Celis_2026}. Under these assumptions, substituting~\eqref{eq:pol_e} and~\eqref{eq:pol_m} into the right-hand sides of~\eqref{eq:GSTC_H} and~\eqref{eq:GSTC_E} and substituting~\eqref{eq:jump_currents_J} and~\eqref{eq:jump_currents_M} into their left-hand sides yield
\begin{equation}
\label{eq:GSTC_3_J}
\begin{aligned}
 &\mathbf{J}_{\Sigma}(\mathbf{r})= -\frac{\mathrm{j} \omega \varepsilon_0}{2}  \hat{\mathbf{n}}(\mathbf{r}) \times \hat{\mathbf{n}}(\mathbf{r}) \times \big(\bar{\bar{\chi}}_{\mathrm{ee}}(\mathbf{r}) \cdot [\mathbf{E}(\mathbf{r}_{+}) + \mathbf{E}(\mathbf{r}_{-}) ]\big) \\
 &-\frac{\mathrm{j}k_0}{2} \hat{\mathbf{n}}(\mathbf{r}) \times \hat{\mathbf{n}}(\mathbf{r}) \times \big(\bar{\bar{\chi}}_{\mathrm{em}}(\mathbf{r})\cdot  [\mathbf{H}(\mathbf{r}_{+})+\mathbf{H}(\mathbf{r}_{-})] \big)
 \end{aligned}
 \end{equation}
 \begin{equation}
 \label{eq:GSTC_3_M}
 \begin{aligned}
 &\mathbf{M}_{\Sigma}(\mathbf{r}) =-\frac{\mathrm{j}k_0}{2}   \hat{\mathbf{n}}(\mathbf{r}) \times \hat{\mathbf{n}}(\mathbf{r}) \times \big(\bar{\bar{\chi}}_{\mathrm{me}}(\mathbf{r}) \cdot[\mathbf{E}(\mathbf{r}_{+})+\mathbf{E}(\mathbf{r}_{-})] \big) \\
 &-\frac{\mathrm{j} \omega \mu_0}{2} \hat{\mathbf{n}}(\mathbf{r}) \times \hat{\mathbf{n}}(\mathbf{r}) \times \big(\bar{\bar{\chi}}_{\mathrm{mm}}(\mathbf{r}) \cdot [\mathbf{H}(\mathbf{r}_{+})+\mathbf{H}(\mathbf{r}_{-})]\big)
 \end{aligned}    
\end{equation}
for $\mathbf{r} \in S$. The total electric and magnetic fields $\mathbf{E}(\mathbf{r})$ and $\mathbf{H}(\mathbf{r})$ are decomposed into incident fields $\mathbf{E}^{\mathrm{inc}}(\mathbf{r})$ and $\mathbf{H}^{\mathrm{inc}}(\mathbf{r})$, and scattered fields $\mathbf{E}^{\mathrm{sca}}(\mathbf{r})$ and $\mathbf{H}^{\mathrm{sca}}(\mathbf{r})$ as
\begin{equation}
\label{eq:total_field_E}
    \mathbf{E}(\mathbf{r}) = \mathbf{E}^{\mathrm{inc}}(\mathbf{r}) + \mathbf{E}^{\mathrm{sca}}(\mathbf{r})
\end{equation}
\begin{equation}
\label{eq:total_field_H}
    \mathbf{H}(\mathbf{r}) = \mathbf{H}^{\mathrm{inc}}(\mathbf{r}) + \mathbf{H}^{\mathrm{sca}}(\mathbf{r}).
\end{equation}
Here, the scattered fields $\mathbf{E}^{\mathrm{sca}}(\mathbf{r})$ and $\mathbf{H}^{\mathrm{sca}}(\mathbf{r})$ are expressed in terms of the equivalent surface currents $\mathbf{J}_{\Sigma}(\mathbf{r})$ and $\mathbf{M}_{\Sigma}(\mathbf{r})$~\cite{jin2015theory, peterson1998}:
\begin{equation}
\label{eq:scat_field_E}
    \mathbf{E}^\mathrm{sca}(\mathbf{r})= \eta_0\mathcal{L}[\mathbf{J}_{\Sigma}](\mathbf{r})-\mathcal{K}[\mathbf{M}_{\Sigma}](\mathbf{r})
\end{equation}
\begin{equation}
\label{eq:scat_field_H}
    \mathbf{H}^\mathrm{sca}(\mathbf{r})=\frac{1}{\eta_0}\mathcal{L}[\mathbf{M}_{\Sigma}](\mathbf{r})+\mathcal{K}[\mathbf{J}_{\Sigma}](\mathbf{r})
\end{equation}
where the surface integral operators $\mathcal{L}[\mathbf{X}](\mathbf{r})$ and $\mathcal{K}[\mathbf{X}](\mathbf{r})$ are 
\begin{equation}
\label{eq:integ_op_L}
\mathcal{L}[\mathbf{X}](\mathbf{r})=  \underbrace{-\mathrm{j}k_0\int_S G(\mathbf{r}, \mathbf{r}^{\prime}) \mathbf{X}(\mathbf{r}^{\prime}) \,ds^{\prime}}_{\displaystyle \mathcal{L}_{\mathrm{A}}[\mathbf{X}](\mathbf{r})}\underbrace{-\frac{\mathrm{j}}{k_0}\nabla \int_S G(\mathbf{r}, \mathbf{r}^{\prime}) \nabla^{\prime} \cdot \mathbf{X}(\mathbf{r}^{\prime}) \,ds^{\prime}}_{\displaystyle \mathcal{L}_{\mathrm{\phi}}[\mathbf{X}](\mathbf{r})}
\end{equation}
\begin{equation}
\label{eq:integ_op_K}
\mathcal{K}[\mathbf{X}](\mathbf{r})=  \int_{S} \nabla  G (\mathbf{r}, \mathbf{r}^{\prime}) \times \mathbf{X}(\mathbf{r}^{\prime})  \,ds^{\prime}
\end{equation}
and $G (\mathbf{r}, \mathbf{r}^{\prime})= e^{-\mathrm{j}k_0|\mathbf{r}-\mathbf{r}^{\prime}|}/(4\pi |\mathbf{r}-\mathbf{r}^{\prime}|)$ is the Green function of the homogeneous background medium. To obtain the SIE-GSTC system, $\mathbf{E}(\mathbf{r}_{+}) + \mathbf{E}(\mathbf{r}_{-})$ and $\mathbf{H}(\mathbf{r}_{+}) + \mathbf{H}(\mathbf{r}_{-})$ on the right-hand sides of~\eqref{eq:GSTC_3_J} and~\eqref{eq:GSTC_3_M} are expressed using~\eqref{eq:total_field_E}-\eqref{eq:total_field_H} and~\eqref{eq:scat_field_E}-\eqref{eq:scat_field_H} as
\begin{equation}
\label{eq:rhs_fields_E}
\begin{aligned}
\mathbf{E}(\mathbf{r}_{+}) + \mathbf{E}(\mathbf{r}_{-})& =\mathbf{E}^{\mathrm{inc}}(\mathbf{r}_{+}) +\mathbf{E}^{\mathrm{inc}}(\mathbf{r}_{-})\\
&+\eta_0\mathcal{L}[\mathbf{J}_{\Sigma}](\mathbf{r}_{+})+\eta_0\mathcal{L}[\mathbf{J}_{\Sigma}](\mathbf{r}_{-})
-\mathcal{K}[\mathbf{M}_{\Sigma}](\mathbf{r}_{+})-\mathcal{K}[\mathbf{M}_{\Sigma}](\mathbf{r}_{-})
\end{aligned}
\end{equation}
\begin{equation}
\label{eq:rhs_fields_H}
\begin{aligned}
\mathbf{H}(\mathbf{r}_{+}) + \mathbf{H}(\mathbf{r}_{-})& =\mathbf{H}^{\mathrm{inc}}(\mathbf{r}_{+}) +\mathbf{H}^{\mathrm{inc}}(\mathbf{r}_{-})\\
&+\frac{1}{\eta_0}\mathcal{L}[\mathbf{M}_{\Sigma}](\mathbf{r}_{+})+\frac{1}{\eta_0}\mathcal{L}[\mathbf{M}_{\Sigma}](\mathbf{r}_{-})+\mathcal{K}[\mathbf{J}_{\Sigma}](\mathbf{r}_{+})+\mathcal{K}[\mathbf{J}_{\Sigma}](\mathbf{r}_{-}).
\end{aligned}
\end{equation}
When substituting~\eqref{eq:rhs_fields_E} and~\eqref{eq:rhs_fields_H} into~\eqref{eq:GSTC_3_J} and~\eqref{eq:GSTC_3_M}, one needs to consider that $\mathbf{r}_{+}$ and $\mathbf{r}_{-}$ approach $\mathbf{r} \in S$. Consequently, the free terms arising from $\mathcal{K}[\mathbf{M}_{\Sigma}](\mathbf{r}_{+})$ 
and $\mathcal{K}[\mathbf{M}_{\Sigma}](\mathbf{r}_{-})$ cancel each other, and similarly those from $\mathcal{K}[\mathbf{J}_{\Sigma}](\mathbf{r}_{+})$ and $\mathcal{K}[\mathbf{J}_{\Sigma}](\mathbf{r}_{-})$ cancel each other, since $\mathbf{r}_{+}$ and $\mathbf{r}_{-}$ approach $\mathbf{r} \in S$ from opposite directions. The resulting SIE-GSTC system reads 
\begin{equation}
\label{eq:SIE-GSTC_J}
\begin{aligned}
 &\mathbf{J}_{\Sigma}(\mathbf{r}) + \mathrm{j} \omega \varepsilon_0   \hat{\mathbf{n}}(\mathbf{r}) \times \hat{\mathbf{n}}(\mathbf{r}) \times \big( \bar{\bar{\chi}}_{\mathrm{ee}}(\mathbf{r}) \cdot [\eta_0\mathcal{L}[\mathbf{J}_{\Sigma}](\mathbf{r})- \tilde{\mathcal{K}}[\mathbf{M}_{\Sigma}](\mathbf{r})]\big)+ \mathrm{j}k_0 \hat{\mathbf{n}}(\mathbf{r}) \times \hat{\mathbf{n}}(\mathbf{r}) \\
 &\times \big(\bar{\bar{\chi}}_{\mathrm{em}}(\mathbf{r})\cdot [\frac{1}{\eta_0}\mathcal{L}[\mathbf{M}_{\Sigma}](\mathbf{r}) + \tilde{\mathcal{K}}[\mathbf{J}_{\Sigma}](\mathbf{r})]\big)= -\mathrm{j} \omega \varepsilon_0  \hat{\mathbf{n}}(\mathbf{r}) \times \hat{\mathbf{n}}(\mathbf{r}) \times \big[\bar{\bar{\chi}}_{\mathrm{ee}}(\mathbf{r}) \cdot\mathbf{E}^{\mathrm{inc}}(\mathbf{r})\big] \\
 &- \mathrm{j}k_0 \hat{\mathbf{n}}(\mathbf{r}) \times \hat{\mathbf{n}}(\mathbf{r}) \times \big[\bar{\bar{\chi}}_{\mathrm{em}}(\mathbf{r})\cdot \mathbf{H}^{\mathrm{inc}}(\mathbf{r})\big]
 \end{aligned}
 \end{equation}
 \begin{equation}
 \label{eq:SIE-GSTC_M}
 \begin{aligned}
 &\mathbf{M}_{\Sigma}(\mathbf{r})+ \mathrm{j}k_0   \hat{\mathbf{n}}(\mathbf{r}) \times \hat{\mathbf{n}}(\mathbf{r}) \times \big(\bar{\bar{\chi}}_{\mathrm{me}}(\mathbf{r}) \cdot [\eta_0\mathcal{L}[\mathbf{J}_{\Sigma}](\mathbf{r})- \tilde{\mathcal{K}}[\mathbf{M}_{\Sigma}](\mathbf{r})]\big)+ \mathrm{j} \omega \mu_0  \hat{\mathbf{n}}(\mathbf{r}) \times \hat{\mathbf{n}}(\mathbf{r})\\
 &\times \big(\bar{\bar{\chi}}_{\mathrm{mm}}(\mathbf{r}) \cdot[\frac{1}{\eta_0}\mathcal{L}[\mathbf{M}_{\Sigma}](\mathbf{r}) + \tilde{\mathcal{K}}[\mathbf{J}_{\Sigma}](\mathbf{r})]\big)= \\
 & -\mathrm{j}k_0   \hat{\mathbf{n}}(\mathbf{r}) \times \hat{\mathbf{n}}(\mathbf{r}) \times\big[\bar{\bar{\chi}}_{\mathrm{me}}(\mathbf{r}) \cdot\mathbf{E}^{\mathrm{inc}}(\mathbf{r})\big] - \mathrm{j} \omega \mu_0   \hat{\mathbf{n}}(\mathbf{r}) \times \hat{\mathbf{n}}(\mathbf{r}) \times \big[\bar{\bar{\chi}}_{\mathrm{mm}}(\mathbf{r}) \cdot\mathbf{H}^{\mathrm{inc}}(\mathbf{r})\big]
 \end{aligned}    
\end{equation}
for $\mathbf{r} \in S$. Here, $\tilde{\mathcal{K}}[\mathbf{X}](\mathbf{r})$ denotes the principal-value part of $\mathcal{K}[\mathbf{X}](\mathbf{r})$, obtained after the cancellation of the free terms described above. The magnetic current density is redefined as $\mathbf{M}_{\Sigma}(\mathbf{r}) \leftarrow  \mathbf{M}_{\Sigma}(\mathbf{r})/\eta_0$~\cite{jin2015theory}, so that $\mathbf{J}_{\Sigma}(\mathbf{r})$ and $\mathbf{M}_{\Sigma}(\mathbf{r})$ have the same units and comparable magnitudes, yielding a well-balanced system. The final form of the SIE-GSTC system reads:
\begin{equation}
\label{eq:SIE-GSTC_2_J}
\begin{aligned}
 &\mathbf{J}_{\Sigma}(\mathbf{r})+ \mathrm{j}k_0 \hat{\mathbf{n}}(\mathbf{r}) \times \hat{\mathbf{n}}(\mathbf{r}) \times \big(\bar{\bar{\chi}}_{\mathrm{ee}}(\mathbf{r}) \cdot [\mathcal{L}[\mathbf{J}_{\Sigma}](\mathbf{r})- \tilde{\mathcal{K}}[\mathbf{M}_{\Sigma}](\mathbf{r})] \big) + \mathrm{j}k_0 \hat{\mathbf{n}}(\mathbf{r}) \times \hat{\mathbf{n}}(\mathbf{r}) \\
 &\times \big(\bar{\bar{\chi}}_{\mathrm{em}}(\mathbf{r})\cdot [\mathcal{L}[\mathbf{M}_{\Sigma}](\mathbf{r}) + \tilde{\mathcal{K}}[\mathbf{J}_{\Sigma}](\mathbf{r})] \big)= -\frac{\mathrm{j}k_0}{\eta_0}  \hat{\mathbf{n}}(\mathbf{r}) \times \hat{\mathbf{n}}(\mathbf{r}) \times \big[\bar{\bar{\chi}}_{\mathrm{ee}}(\mathbf{r}) \cdot \mathbf{E}^{\mathrm{inc}}(\mathbf{r})\big] \\
 &- \mathrm{j}k_0 \hat{\mathbf{n}}(\mathbf{r}) \times \hat{\mathbf{n}}(\mathbf{r}) \times \big[\bar{\bar{\chi}}_{\mathrm{em}}(\mathbf{r})\cdot\mathbf{H}^{\mathrm{inc}}(\mathbf{r})\big]
 \end{aligned}
 \end{equation}
 \begin{equation}
 \label{eq:SIE-GSTC_2_M}
 \begin{aligned}
 &\eta_0 \mathbf{M}_{\Sigma}(\mathbf{r}) + \mathrm{j}k_0\eta_0 \hat{\mathbf{n}}(\mathbf{r}) \times \hat{\mathbf{n}}(\mathbf{r}) \times \big(\bar{\bar{\chi}}_{\mathrm{me}}(\mathbf{r}) \cdot [\mathcal{L}[\mathbf{J}_{\Sigma}](\mathbf{r})- \tilde{\mathcal{K}}[\mathbf{M}_{\Sigma}](\mathbf{r})] \big)+ \mathrm{j}k_0\eta_0 \hat{\mathbf{n}}(\mathbf{r}) \times \hat{\mathbf{n}}(\mathbf{r}) \\
 &\times \big(\bar{\bar{\chi}}_{\mathrm{mm}}(\mathbf{r}) \cdot [\mathcal{L}[\mathbf{M}_{\Sigma}](\mathbf{r}) + \tilde{\mathcal{K}}[\mathbf{J}_{\Sigma}](\mathbf{r})] \big) = -\mathrm{j}k_0 \hat{\mathbf{n}}(\mathbf{r}) \times \hat{\mathbf{n}}(\mathbf{r}) \times \big[\bar{\bar{\chi}}_{\mathrm{me}}(\mathbf{r}) \cdot\mathbf{E}^{\mathrm{inc}}(\mathbf{r})\big] \\
 &- \mathrm{j}k_0\eta_0  \hat{\mathbf{n}}(\mathbf{r}) \times \hat{\mathbf{n}}(\mathbf{r}) \times \big[\bar{\bar{\chi}}_{\mathrm{mm}}(\mathbf{r}) \cdot \mathbf{H}^{\mathrm{inc}}(\mathbf{r})\big]
 \end{aligned}    
\end{equation}
for $\mathbf{r}\in S$. The SIE-GSTC system~\eqref{eq:SIE-GSTC_2_J} and~\eqref{eq:SIE-GSTC_2_M} is the starting point for the numerical discretization described in Section~\ref{sec:discretization}.

\subsection{Discretization}\label{sec:discretization}
To numerically solve the SIE-GSTC system~\eqref{eq:SIE-GSTC_2_J}--\eqref{eq:SIE-GSTC_2_M}, $S$ is discretized into a triangular mesh, and the unknowns $\mathbf{J}_{\Sigma}(\mathbf{r})$ and $\mathbf{M}_{\Sigma}(\mathbf{r})$ are expanded as
\begin{equation}
\label{eq:basis_exp_J}
    \mathbf{J}_{\Sigma}(\mathbf{r}) = \sum_{n=1}^{N} \{\bar{J}\}_n \mathbf{f}_n(\mathbf{r})
\end{equation}
\begin{equation}
\label{eq:basis_exp_M}
    \mathbf{M}_{\Sigma}(\mathbf{r}) = \sum_{n=1}^{N} \{\bar{M}\}_n \mathbf{f}_n(\mathbf{r})
\end{equation}
where $\mathbf{f}_n(\mathbf{r})$ denotes the RWG basis function associated with interior edge $n$ and defined over its two adjacent triangular patches, and $\bar{J}$ and $\bar{M}$ denote the vectors of unknown expansion coefficients for $\mathbf{J}_{\Sigma}(\mathbf{r})$ and $\mathbf{M}_{\Sigma}(\mathbf{r})$, respectively. Here, $N$ is the total number of interior edges in the mesh. The absence of basis functions on boundary edges implicitly enforces the vanishing-current condition at the rim of the open metasurface. 

Inserting the expansions~\eqref{eq:basis_exp_J}-\eqref{eq:basis_exp_M} into the SIE-GSTC system~\eqref{eq:SIE-GSTC_2_J}--\eqref{eq:SIE-GSTC_2_M} and applying Galerkin testing with testing function $\mathbf{f}_m(\mathbf{r})$ for $m = 1, 2, \dots, N$ yield a linear system of dimension $2N \times 2N$, given by:
\begin{align}
\underbrace{{\begin{bmatrix}
\bar{\bar{Z}}_{\mathrm{JJ}}&\bar{\bar{Z}}_{\mathrm{JM}}\\
\bar{\bar{Z}}_{\mathrm{MJ}}&\bar{\bar{Z}}_{\mathrm{MM}}
\end{bmatrix}}}_{\displaystyle \bar{\bar{Z}}}\cdot
\underbrace{{\begin{bmatrix}
\bar{J}\\
\bar{M}\\
\end{bmatrix}}}_{\displaystyle \bar{I}}=
\underbrace{{\begin{bmatrix}
\bar{V}_{\mathrm{J}}^{\mathrm{inc}}\\
\bar{V}_{\mathrm{M}}^{\mathrm{inc}}\\
\end{bmatrix}}}_{\displaystyle \bar{V}^{\mathrm{inc}}}.\label{eq:mom_equation}
\end{align}
Here, $\bar{\bar{Z}}_{\mathrm{JJ}}$, $\bar{\bar{Z}}_{\mathrm{JM}}$, $\bar{\bar{Z}}_{\mathrm{MJ}}$, and $\bar{\bar{Z}}_{\mathrm{MM}}$ are block matrices of dimensions $N \times N$, and $\bar{V}_{\mathrm{J}}^{\mathrm{inc}}$ and $\bar{V}_{\mathrm{M}}^{\mathrm{inc}}$ are vectors of dimension $N$ that contain the tested incident fields. The entries of $\bar{\bar{Z}}_{\mathrm{JJ}}$, $\bar{\bar{Z}}_{\mathrm{JM}}$, $\bar{\bar{Z}}_{\mathrm{MJ}}$, and $\bar{\bar{Z}}_{\mathrm{MM}}$ are
\begin{equation}
\label{eq:Z_JJ}
    \{\bar{\bar{Z}}_{\mathrm{JJ}}\}_{mn}=\langle\mathbf{f}_m(\mathbf{r}),\mathbf{f}_n(\mathbf{r})\rangle
    -\mathrm{j}k_0\langle \bar{\bar{\chi}}^{\mathsf{T}}_{\mathrm{ee}}(\mathbf{r}) \cdot \mathbf{f}_m(\mathbf{r}), \mathcal{L}[\mathbf{f}_{n}](\mathbf{r})\rangle-\mathrm{j}k_0 \langle \bar{\bar{\chi}}^{\mathsf{T}}_{\mathrm{em}}(\mathbf{r}) \cdot \mathbf{f}_m(\mathbf{r}), \tilde{\mathcal{K}} [\mathbf{f}_{n}](\mathbf{r})\rangle
\end{equation}
\begin{equation}
\label{eq:Z_JM}
    \{\bar{\bar{Z}}_{\mathrm{JM}}\}_{mn}=\mathrm{j} k_0\langle \bar{\bar{\chi}}^{\mathsf{T}}_{\mathrm{ee}}(\mathbf{r}) \cdot \mathbf{f}_m(\mathbf{r}), \tilde{\mathcal{K}}[\mathbf{f}_{n}](\mathbf{r})\rangle-\mathrm{j} k_0\langle \bar{\bar{\chi}}^{\mathsf{T}}_{\mathrm{em}}(\mathbf{r}) \cdot\mathbf{f}_m(\mathbf{r}), \mathcal{L}[\mathbf{f}_{n}](\mathbf{r})\rangle
\end{equation}
\begin{equation}
\label{eq:Z_MJ}
    \{\bar{\bar{Z}}_{\mathrm{MJ}}\}_{mn}=-\mathrm{j}  k_0 \eta_0\langle \bar{\bar{\chi}}^{\mathsf{T}}_{\mathrm{me}}(\mathbf{r}) \cdot \mathbf{f}_m(\mathbf{r}), \mathcal{L}[\mathbf{f}_{n}](\mathbf{r})\rangle
    -\mathrm{j} k_0 \eta_0\langle \bar{\bar{\chi}}^{\mathsf{T}}_{\mathrm{mm}}(\mathbf{r}) \cdot \mathbf{f}_m(\mathbf{r}), \tilde{\mathcal{K}}[\mathbf{f}_{n}](\mathbf{r})\rangle
\end{equation}
\begin{equation}
\label{eq:Z_MM}
\begin{aligned}
     \{\bar{\bar{Z}}_{\mathrm{MM}}\}_{mn}&=\eta_{0}\langle\mathbf{f}_m(\mathbf{r}),\mathbf{f}_n(\mathbf{r})\rangle+\mathrm{j} k_0  \eta_{0} \langle \bar{\bar{\chi}}^{\mathsf{T}}_{\mathrm{me}}(\mathbf{r}) \cdot \mathbf{f}_m(\mathbf{r}), \tilde{\mathcal{K}}[\mathbf{f}_{n}](\mathbf{r})\rangle\\
&-\mathrm{j} k_0 \eta_0\langle \bar{\bar{\chi}}^{\mathsf{T}}_{\mathrm{mm}}(\mathbf{r}) \cdot \mathbf{f}_m(\mathbf{r}), \mathcal{L}[\mathbf{f}_{n}](\mathbf{r})\rangle
\end{aligned}
\end{equation}
for $m, n =1,2,\ldots,N$. The entries of $\bar{V}_{\mathrm{J}}^{\mathrm{inc}}$ and $\bar{V}_{\mathrm{M}}^{\mathrm{inc}}$ are
\begin{equation}
\label{eq:Einc}
    \{\bar{V}_{\mathrm{J}}^{\mathrm{inc}}\}_m=\mathrm{j} \frac{k_0}{\eta_0}\langle \bar{\bar{\chi}}^{\mathsf{T}}_{\mathrm{ee}}(\mathbf{r}) \cdot \mathbf{f}_m(\mathbf{r}),  \mathbf{E}^{\mathrm{inc}}(\mathbf{r}) \rangle+ \mathrm{j}k_0\langle \bar{\bar{\chi}}^{\mathsf{T}}_{\mathrm{em}}(\mathbf{r}) \cdot \mathbf{f}_m(\mathbf{r}), \mathbf{H}^{\mathrm{inc}}(\mathbf{r}) \rangle
\end{equation}
\begin{equation}
\label{eq:Hinc}
    \{\bar{V}_{\mathrm{M}}^{\mathrm{inc}}\}_m=\mathrm{j}k_0\langle \bar{\bar{\chi}}^{\mathsf{T}}_{\mathrm{me}}(\mathbf{r}) \cdot \mathbf{f}_m(\mathbf{r}), \mathbf{E}^{\mathrm{inc}}(\mathbf{r}) \rangle+\mathrm{j} k_0 \eta_0 \langle \bar{\bar{\chi}}^{\mathsf{T}}_{\mathrm{mm}}(\mathbf{r}) \cdot \mathbf{f}_m(\mathbf{r}), \mathbf{H}^{\mathrm{inc}}(\mathbf{r}) \rangle
\end{equation}
for $m=1,2,\dots,N$. Here, the inner product between two vectors $\mathbf{X}(\mathbf{r})$ and $\mathbf{Y}(\mathbf{r})$ is defined as
\begin{equation}
\langle \mathbf{X}(\mathbf{r}), \mathbf{Y}(\mathbf{r}) \rangle = \int_{S_\mathrm{x} \cap S_\mathrm{y}} \mathbf{X}(\mathbf{r}) \cdot \mathbf{Y}(\mathbf{r})\, ds
\end{equation}
where $S_{\mathrm{x}}$ and $S_{\mathrm{y}}$ denote the supports of $\mathbf{X}(\mathbf{r})$ and $\mathbf{Y}(\mathbf{r})$, respectively. In the derivation of~\eqref{eq:Z_JJ}--\eqref{eq:Hinc}, two identities 
are used. The first simplifies the tangential projection operator:
\begin{equation}
\langle \mathbf{f}_m(\mathbf{r}), \hat{\mathbf{n}}(\mathbf{r})\times 
\hat{\mathbf{n}}(\mathbf{r})\times \mathbf{Y}(\mathbf{r}) \rangle = 
-\langle \mathbf{f}_m(\mathbf{r}),\mathbf{Y}(\mathbf{r})\rangle
\end{equation}
which follows from the ``BAC-CAB'' rule and the fact that $\mathbf{f}_m(\mathbf{r})\cdot\hat{\mathbf{n}}(\mathbf{r}) = 0$, since RWG basis functions are tangential to $S$. The second is the adjoint property of the inner product:
\begin{equation}
\langle \mathbf{X}(\mathbf{r}), \bar{\bar{\chi}}_{\mathrm{ab}}(\mathbf{r})
\cdot \mathbf{Y}(\mathbf{r}) \rangle = \langle 
\bar{\bar{\chi}}^{\mathsf{T}}_{\mathrm{ab}}(\mathbf{r})\cdot \mathbf{X}(\mathbf{r}), 
\mathbf{Y}(\mathbf{r}) \rangle
\end{equation}
where the superscript $\mathsf{T}$ denotes the transpose.

The block matrix entries in~\eqref{eq:Z_JJ}--\eqref{eq:Z_MM} require the computation of surface integrals present in the inner products
\begin{equation}
\label{eq:integ1}
I_{1} = \langle \bar{\bar{\chi}}^{\mathsf{T}}_{\mathrm{ab}}(\mathbf{r}) \cdot \mathbf{f}_m(\mathbf{r}), \tilde{\mathcal{K}}[\mathbf{f}_{n}](\mathbf{r}) \rangle
\end{equation}
\begin{equation}
\label{eq:integ2}
I_{2} = \langle \bar{\bar{\chi}}^{\mathsf{T}}_{\mathrm{ab}} (\mathbf{r}) \cdot \mathbf{f}_m(\mathbf{r}), \mathcal{L}[\mathbf{f}_{n}](\mathbf{r}) \rangle.
\end{equation}
Let $T^{\beta}_n$, $\beta \in \{+,-\}$ represent the two triangular patches adjacent to edge $n$, which is associated with the RWG function $\mathbf{f}_n(\mathbf{r})$. For the computation of $I_1$ and $I_2$, it is assumed that
\begin{equation}
\bar{\bar{\chi}}_{\mathrm{ab}}(\mathbf{r})=\bar{\bar{\chi}}_{\mathrm{ab}}(\mathbf{r}^{\beta}_{\mathrm{c},n})=\bar{\bar{\chi}}_{\mathrm{ab},n}^{\beta}, \;\mathbf{r} \in T^{\beta}_n
\end{equation}
where $\mathbf{r}^{\beta}_{\mathrm{c},n}$ is the center of $T^{\beta}_n$. $I_{1}$ is computed using the Gaussian quadrature rules for triangles~\cite{dunavant1985, cowper1973} together with the standard singularity treatment techniques in~\cite{CFIE_Pasi,hodges_evaluation_1997}. $I_2$ is decomposed into its vector and scalar potential components:
\begin{equation}
\label{eq:L_decomp}
    I_{2} = \underbrace{\langle \bar{\bar{\chi}}^{\mathsf{T}}_{\mathrm{ab}}(\mathbf{r}) \cdot \mathbf{f}_m(\mathbf{r}), \mathcal{L}_{\mathrm{A}}[\mathbf{f}_{n}](\mathbf{r})\rangle}_{\displaystyle I_{\mathrm{A}}} + \underbrace{\langle \bar{\bar{\chi}}^{\mathsf{T}}_{\mathrm{ab}}(\mathbf{r}) \cdot \mathbf{f}_m(\mathbf{r}), \mathcal{L}_{\mathrm{\phi}}[\mathbf{f}_{n}](\mathbf{r})\rangle}_{\displaystyle I_{\phi}}.
\end{equation}
 $\mathcal{L}_{\mathrm{A}}[\mathbf{X}](\mathbf{r})$ and $\mathcal{L}_{\mathrm{\phi}}[\mathbf{X}](\mathbf{r})$ are the integral operators defined in~\eqref{eq:integ_op_L}.  $I_{\mathrm{A}}$ is computed using the Gaussian quadrature rules for triangles~\cite{dunavant1985, cowper1973} together with the  standard singularity treatment techniques in~\cite{wilton_potential_1984, graglia_numerical_1993}. On the other hand, the computation of $I_{\mathrm{\phi}}$ requires more careful treatment due to the presence of $\bar{\bar{\chi}}^{\mathsf{T}}_{\mathrm{ab}}(\mathbf{r})$ in the testing integral as described next.  $I_{\mathrm{\phi}}$  has four contributions coming from the surface integrals defined over combinations of $T^{\alpha}_m$ and $T^{\beta}_n$ with $\alpha, \beta \in \{+,-\}$. Each of these contributions is given by
\begin{equation}
\label{eq:Lphi}
 \begin{aligned}
    \int_{T^{\alpha}_m}  [\bar{\bar{\chi}}_{\mathrm{ab},m}^{\alpha\mathsf{T}} \cdot \mathbf{f}_m(\mathbf{r})] \cdot \nabla \int_{T^{\beta}_n} G(\mathbf{r}, \mathbf{r}^{\prime}) \nabla^{\prime} \cdot \mathbf{f}_n(\mathbf{r}^{\prime})\,ds^{\prime}\,ds.
\end{aligned}    
\end{equation}
where the constant $-\mathrm{j}/k_0$ from~\eqref{eq:integ_op_L} is omitted for brevity. By applying the product rule for the surface divergence and the surface divergence theorem, and using the facts that $\mathbf{f}_n(\mathbf{r})\cdot\hat{\mathbf{n}}(\mathbf{r}) = 0$ and that, by \eqref{eq:chi_tangential}, $\bar{\bar{\chi}}_{\mathrm{ab},m}^{\alpha\mathsf{T}} \cdot \mathbf{f}_m(\mathbf{r})$ is
tangential to $S$,~\eqref{eq:Lphi} is converted into~\cite{CFIE_Pasi}:
\begin{equation}
\label{eq:lphi_d}
\begin{aligned}
&-\int_{T^{\alpha}_m} \nabla_{\mathrm{s}} \cdot [ \bar{\bar{\chi}}_{\mathrm{ab},m}^{\alpha\mathsf{T}} \cdot \mathbf{f}_m(\mathbf{r})] 
\int_{T^{\beta}_n} G(\mathbf{r}, \mathbf{r}^{\prime})\,\nabla_{\mathrm{s}}^{\prime} \cdot \mathbf{f}_n(\mathbf{r}^{\prime})\, ds^{\prime}\, ds\\
&+\int_{\partial T^{\alpha}_m}\hat{\mathbf{m}}(\mathbf{r}) \cdot [ \bar{\bar{\chi}}_{\mathrm{ab},m}^{\alpha\mathsf{T}} \cdot \mathbf{f}_m(\mathbf{r})] 
\int_{T^{\beta}_n} G(\mathbf{r}, \mathbf{r}^{\prime})\,\nabla_{\mathrm{s}}^{\prime} \cdot \mathbf{f}_n(\mathbf{r}^{\prime})\, ds^{\prime}\, dl.    
\end{aligned}
\end{equation}
Here, $\nabla_{\mathrm{s}} \cdot \mathbf{X}(\mathbf{r})=\nabla \cdot \mathbf{X}(\mathbf{r})-\hat{\mathbf{n}}(\mathbf{r}) \cdot[\hat{\mathbf{n}}(\mathbf{r}) \cdot \nabla] \mathbf{X}(\mathbf{r})$ is the surface divergence, $\partial T^{\alpha}_m$ denotes the boundary of $T^{\alpha}_m$, $\hat{\mathbf{m}}(\mathbf{r}) = \hat{\mathbf{t}}(\mathbf{r})\times \hat{\mathbf{n}}(\mathbf{r})$ is the outward-pointing unit conormal vector, with $\hat{\mathbf{t}}(\mathbf{r})$ the unit tangent to $\partial T^{\alpha}_m$. The conormal vector lies in the plane of $T^{\alpha}_m$ and is normal to $\partial T^{\alpha}_m$. 

In contrast to the standard discretization of electric field integral equation (EFIE) using RWG functions, the line integral in~\eqref{eq:lphi_d} does not vanish. In the formulation here, the weighted testing function $\bar{\bar{\chi}}_{\mathrm{ab},m}^{\alpha\mathsf{T}} \cdot \mathbf{f}_m(\mathbf{r})$ is in general rotated within the tangent plane, therefore its conormal component does not vanish on the two non-defining edges of $T^{\alpha}_m$. Moreover, since $\bar{\bar{\chi}}_{\mathrm{ab},m}^{+} \neq \bar{\bar{\chi}}_{\mathrm{ab},m}^{-}$ in general, the contributions from $T^{+}_m$ and $T^{-}_m$ on the shared edge $m$ do not cancel. The line integral therefore accounts for the line charges introduced by the weighting.

Since $\hat{\mathbf{n}}(\mathbf{r}) \cdot\bar{\bar{\chi}}_{\mathrm{ab},m}^{\alpha} \cdot \hat{\mathbf{n}}(\mathbf{r}) = 0$ [see~\eqref{eq:chi_tangential}], the surface divergence in~\eqref{eq:lphi_d} admits the closed form~\cite{Shanker_Anisotropic}:
\begin{equation}
\label{eq:Divs}
\nabla_{\mathrm{s}} \cdot [ \bar{\bar{\chi}}_{\mathrm{ab},m}^{\alpha\mathsf{T}} \cdot \mathbf{f}_m(\mathbf{r}) ]
= \frac{\alpha l_{m}}{2 A_{m}^{\alpha}}\,\operatorname{tr}(\bar{\bar{\chi}}^{\alpha}_{\mathrm{ab},m}),
\; \mathbf{r}\in T_m^{\alpha}
\end{equation}
where $\operatorname{tr}(\bar{\bar{\chi}})$ is the sum of the diagonal elements of $\bar{\bar{\chi}}$, $l_m$ is the length of edge $m$ and $A_m^\alpha$ is the area of $T_m^\alpha$. The inner integral over $T^{\beta}_n$ is common to both terms of \eqref{eq:lphi_d} and is
evaluated using Gaussian quadrature rules for triangles~\cite{dunavant1985, cowper1973} together with the standard singularity treatment techniques in~\cite{wilton_potential_1984, graglia_numerical_1993}, which remain valid when the observation point $\mathbf{r}$ lies on $\partial T^{\alpha}_m$. The outer integration is performed with Gaussian quadrature rules for triangles over $T^{\alpha}_m$ in the first term, and with Gauss--Legendre quadrature~\cite{davis_rabinowitz} along each edge of $\partial T^{\alpha}_m$ in the second term.

\section{Numerical Results}\label{sec:Numerical_Examples}

This section presents numerical examples that evaluate the accuracy and robustness of the proposed SIE-GSTC solver and demonstrate its applicability to open metasurface configurations. For each example, the equivalent susceptibility tensors are synthesized from the prescribed electromagnetic transformation following~\cite{Synthesis_Spherical, Susceptibility_Tensors, achouri2018, liu2019generalized}, with the number of independent tensor components determined by the transformation itself. 
 
In all examples, the metasurface is illuminated by a Gaussian beam propagating along the $z$ direction. The incident electric field is 
\begin{equation}
\label{eq:GaussInc}
\mathbf{E}^{\mathrm{inc}}(\mathbf{r}) = \hat{\mathbf{p}} E_0 \frac{w_\mathrm{b}}{w(d)} e^{-\left[\frac{\rho^2}{w^2(d)}+\mathrm{j}\Phi(d,\rho)\right]}
\end{equation}
where $E_0 = 1\,\mathrm{V/m}$ is the amplitude, $\hat{\mathbf{p}} \in \{\hat{\mathbf{x}},\hat{\mathbf{y}}\}$ is the polarization, $d = \mathbf{r}\cdot\hat{\mathbf{z}}$ is the axial distance along the propagation direction, $\rho = \|\mathbf{r}-d\hat{\mathbf{z}}\|$ is the transverse radial distance, and $w_{\mathrm{b}}$ is the beam radius at $d=0$. The beam radius $w(d)$ and the phase variation $\Phi(d,\rho)$ are defined as
\begin{equation}
\begin{aligned}
w(d) &= w_\mathrm{b}\sqrt{1+\left(\frac{d}{d_\mathrm{R}}\right)^2} \\
\Phi(d,\rho) &= k_0 d + \frac{k_0 \rho^2}{2R(d)} - \Psi(d)
\end{aligned}
\end{equation}
where $R(d) = d[1+(d_\mathrm{R}/d)^2]$ is the radius of curvature, $\Psi(d) = \tan^{-1}(d/d_\mathrm{R})$ is the Gouy phase shift, and $d_\mathrm{R} = {\pi w_\mathrm{b}^2}/{\lambda_0}$ is the Rayleigh range. 

Metasurfaces reside in free space, and accordingly $\varepsilon_0$, $\mu_0$, $\eta_0$, $k_0$, and $\lambda_0$ denote the free-space permittivity, permeability, wave impedance, wavenumber, and wavelength, respectively. 

To quantify the accuracy of the proposed SIE-GSTC solver, the relative $\ell_2$-norm error of the scattered fields is evaluated at observation points $\mathbf{r}_{p}$, $p=1,2,\ldots,N_{\mathrm{p}}$ as
\begin{equation}
\label{eq:error_metric}
\mathrm{err}_{\ell_2} = \sqrt{ \frac{ \sum_{p=1}^{N_{\mathrm{p}}} \left\| \mathbf{E}_{\mathrm{sim}}(\mathbf{r}_p) - \mathbf{E}_{\mathrm{ref}}(\mathbf{r}_p) \right\|^2 }{ \sum_{p=1}^{N_{\mathrm{p}}} \left\| \mathbf{E}_{\mathrm{ref}}(\mathbf{r}_p) \right\|^2 } }.
\end{equation}
Here, $\mathbf{E}_{\mathrm{sim}}(\mathbf{r})$ denotes the scattered electric field computed using the SIE-GSTC solver and $\mathbf{E}_{\mathrm{ref}}(\mathbf{r})$ denotes the reference scattered electric field.
 
For all examples, the matrix system in~\eqref{eq:mom_equation} is solved iteratively with diagonal preconditioning, using either the transpose-free quasi-minimal residual (TFQMR) method~\cite{TFQMR2} or the generalized minimal residual (GMRES) method~\cite{GMRES}. The specific solver is indicated in each example. The iterations are terminated when the relative residual falls below the specified tolerance $\epsilon$:
\begin{equation}
\label{eq:conv}
\frac{\left\| \bar{V}^{\mathrm{inc}} - \bar{\bar{Z}}\bar{I}^{(n)} \right\|}{\left\| \bar{V}^{\mathrm{inc}} \right\|} < \epsilon = 10^{-3}.
\end{equation}
Here, $\bar{I}^{(n)}$ is the iterate at iteration $n$.
\subsection{Polarization Rotator}\label{sec:pol_rot}
The metasurface considered in this section is designed to rotate the electric field of an $x$-polarized plane wave propagating along the $z$ direction by an angle $\theta$ upon transmission. The prescribed field transformation specifies fields on the incident side, $\mathbf{E}(\mathbf{r}_{-})$, and the transmitted side, $\mathbf{E}(\mathbf{r}_{+})$, as
\begin{equation}
\label{eq:transform_rot}
\begin{aligned}
\mathbf{E}(\mathbf{r}_{-}) &= \hat{\mathbf{x}} e^{-\mathrm{j}k_0 z}\\
\mathbf{E}(\mathbf{r}_{+}) &= \hat{\mathbf{x}}\cos\theta\, e^{-\mathrm{j}k_0 z} 
+ \hat{\mathbf{y}} \sin\theta\, e^{-\mathrm{j}k_0 z}.
\end{aligned}
\end{equation}
Polarization rotation can be realized either through strictly monoanisotropic transformations, for which 
$\bar{\bar{\chi}}_{\mathrm{em}}(\mathbf{r}) = \bar{\bar{\chi}}_{\mathrm{me}}(\mathbf{r}) = \bar{\bar{0}}$, or through bianisotropic transformations, for which $\bar{\bar{\chi}}_{\mathrm{ee}}(\mathbf{r}) = \bar{\bar{\chi}}_{\mathrm{mm}}(\mathbf{r}) = \bar{\bar{0}}$, as demonstrated in~\cite{achouri2018,journal1_Celis_2026}. 

For the monoanisotropic case, the required susceptibility tensors exhibit gyrotropic behavior through asymmetric off-diagonal components in $\bar{\bar{\chi}}_{\mathrm{ee}}$ and $\bar{\bar{\chi}}_{\mathrm{mm}}$~\cite{MoM_Sandeep,Synthesis_Spherical,Susceptibility_Tensors}. The nonzero components of the resulting tensors are
\begin{equation}
\begin{aligned}
     \chi_{\mathrm{ee}}^{xy}(\mathbf{r})&=-{2\mathrm{j}(1-\cos{\theta})}/({k_0\sin{\theta}})\\
     \chi_{\mathrm{ee}}^{yx}(\mathbf{r})&={2\mathrm{j}\sin{\theta}}/[{k_0(\cos{\theta}+1)}]\\
       \chi_{\mathrm{mm}}^{xy}(\mathbf{r})&={-2\mathrm{j}\sin{\theta}}/[{ k_0(\cos{\theta}+1)}]\\
     \chi_{\mathrm{mm}}^{yx}(\mathbf{r})&=-{2\mathrm{j}(\cos{\theta}-1)}/({k_0\sin{\theta}}).
\end{aligned}    
\end{equation}
The asymmetry of $\bar{\bar{\chi}}_{\mathrm{ee}}(\mathbf{r})$ and $\bar{\bar{\chi}}_{\mathrm{mm}}(\mathbf{r})$ renders this realization nonreciprocal. For the bianisotropic case, reciprocal rotation is achieved through coupling tensors satisfying the reciprocity condition $\bar{\bar{\chi}}_{\mathrm{me}}(\mathbf{r}) = -\bar{\bar{\chi}}_{\mathrm{em}}^{\mathsf{T}}(\mathbf{r})$~\cite{Lorentz_Reciprocity}. The nonzero components of the resulting tensors are
\begin{equation}
 \begin{aligned}
     \chi_{\mathrm{em}}^{xx}(\mathbf{r})&=-{2\mathrm{j}(\cos{\theta}-1)}/({k_0\sin{\theta}})\\
     \chi_{\mathrm{em}}^{yy}(\mathbf{r})&={2\mathrm{j}\sin{\theta}}/[{k_0(\cos{\theta}+1)}]\\
     \chi_{\mathrm{me}}^{xx}(\mathbf{r})&=-{2\mathrm{j}\sin{\theta}}/[{k_0 (\cos{\theta}+1)}]\\
     \chi_{\mathrm{me}}^{yy}(\mathbf{r})&={2\mathrm{j}(\cos{\theta}-1)}/({k_0 \sin{\theta}}).
\end{aligned}   
\end{equation}
The susceptibility tensors in both cases are synthesized for a flat metasurface in the $xy$-plane, with their components expressed in global Cartesian coordinates. For the curved metasurface considered later in this section, these tensors are imposed in the local tangent frame of each triangular patch of the mesh.

The SIE-GSTC solver is validated by applying the synthesized susceptibility tensors to both a planar and a curved metasurface, described in the following two subsections. 

\subsubsection{Planar Metasurface}\label{sec:pol_rot_flat}
For the planar metasurface, $S$ is a circular disk in the $xy$-plane, centered at $(0, 0, 0.4\lambda_0)$, with a radius of $R_{\mathrm{d}}=4\lambda_0$. Both the monoanisotropic and bianisotropic realizations of the rotator are simulated for three rotation angles,  $\theta \in \{30^{\circ}, 45^{\circ}, 60^{\circ}\}$. Surface $S$ is discretized into a triangular mesh with an average edge length of approximately $\lambda_0/10$, yielding a matrix of dimension $35\,100 \times 35\,100$. It is illuminated by the Gaussian beam in~\eqref{eq:GaussInc} with polarization $\hat{\mathbf{p}}=\hat{\mathbf{x}}$ and the beam radius $w_{\mathrm{b}}=2\lambda_0$ at an excitation frequency of $f = 60\,\mathrm{GHz}$. The matrix systems for both realizations are solved using TFQMR. 

\begin{figure}[!ht]
\centering
\subfloat[]{\includegraphics[width=0.46\linewidth]{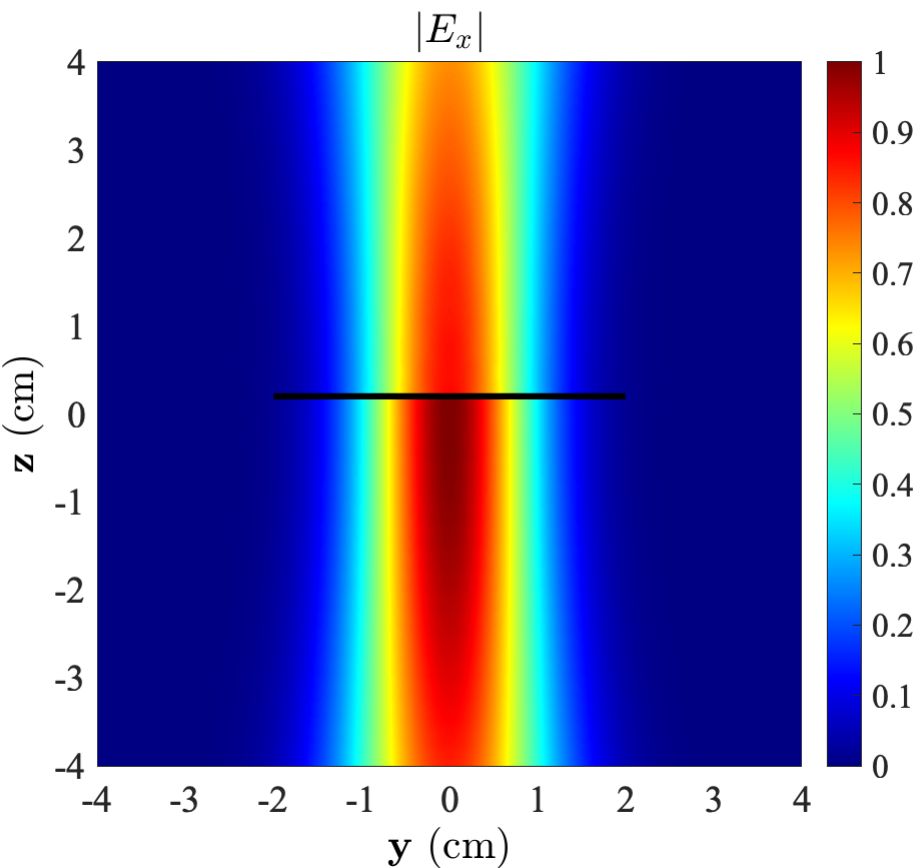}}\hspace{12pt}
\subfloat[]{\includegraphics[width=0.46\linewidth]{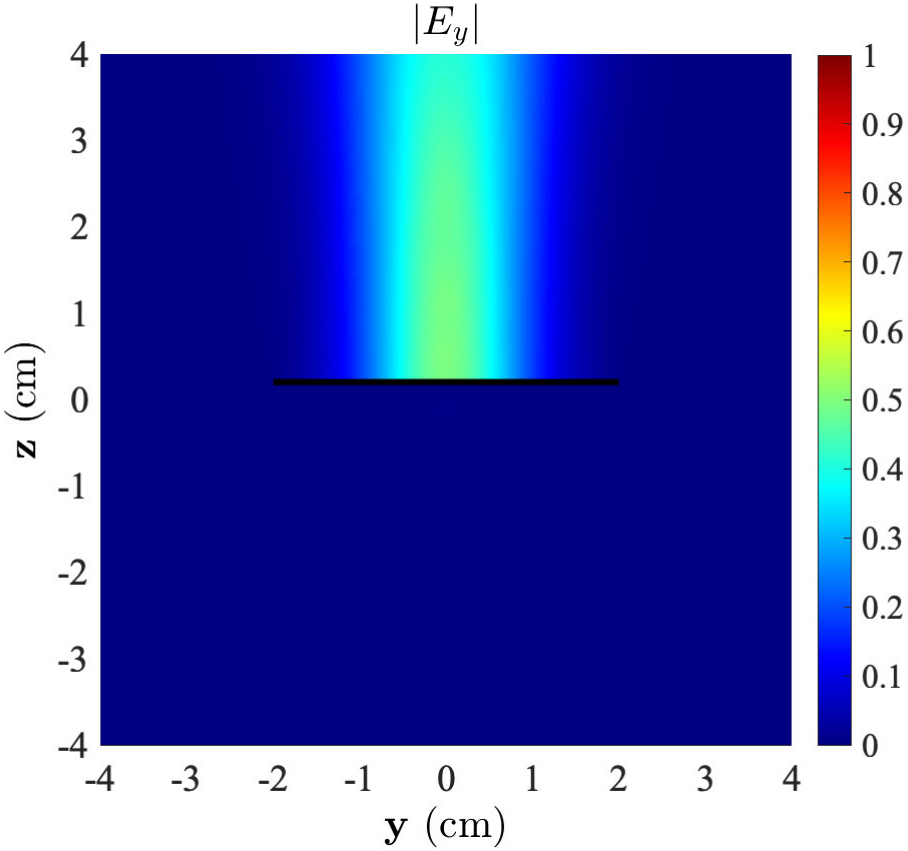}}\vspace{12pt}\\
\subfloat[]{\includegraphics[width=0.6\linewidth]{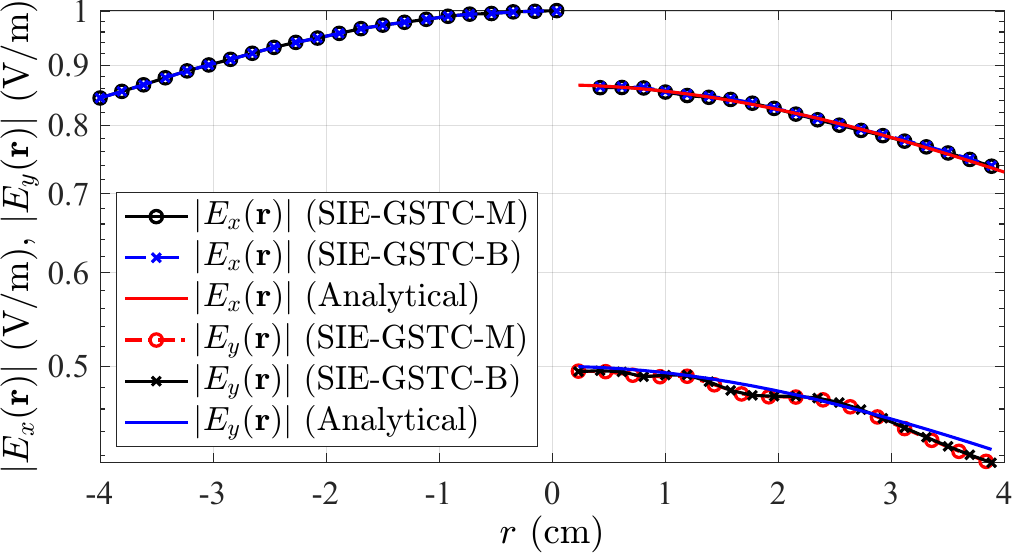}}
\caption{Polarization rotator with $\theta=30^{\circ}$. (a) $|E_x(\mathbf{r})|$ and (b) $|E_y(\mathbf{r})|$ computed using the SIE-GSTC solver on the $yz$-plane for the monoanisotropic realization. (c) $|E_x(\mathbf{r})|$ and $|E_y(\mathbf{r})|$ computed using the SIE-GSTC solver along the $z$-axis, $z \in [-4, 4]\, \mathrm{cm}$, for both the monoanisotropic (``M'') and bianisotropic (``B'') realizations, compared against the analytical expressions.}
\label{Fig:Pol_Rot30}
\end{figure}

\begin{figure}[!ht]
\centering
\subfloat[]{\includegraphics[width=0.46\linewidth]{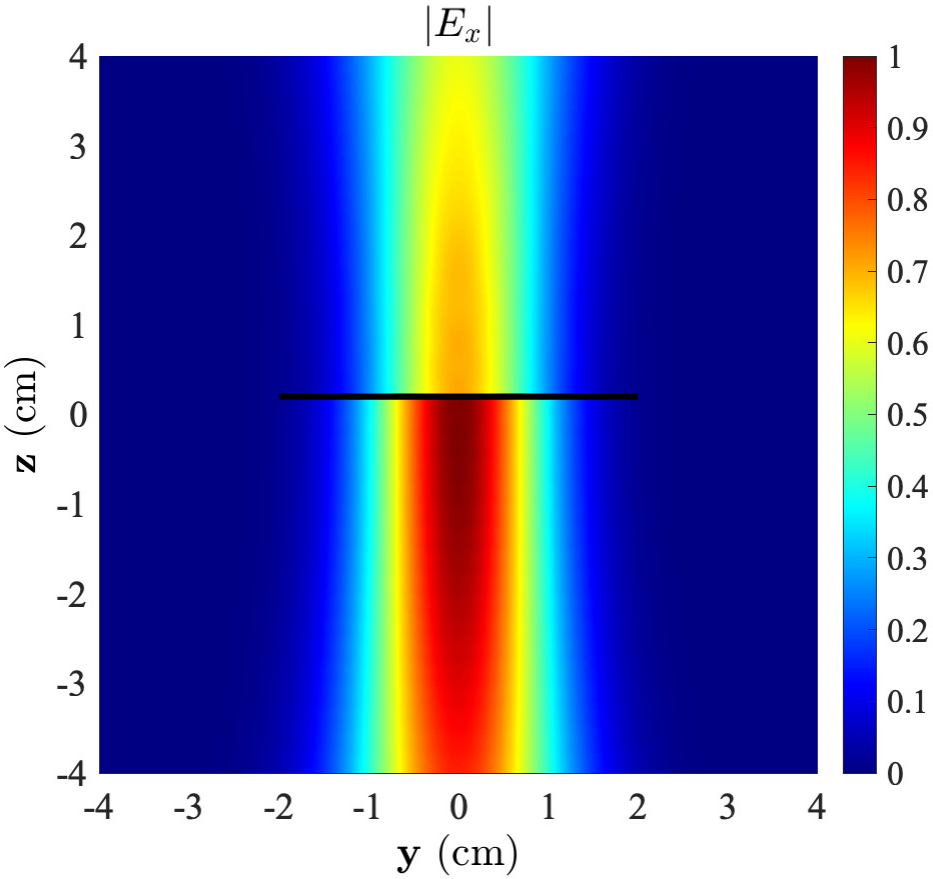}}\hspace{12pt}
\subfloat[]{\includegraphics[width=0.46\linewidth]{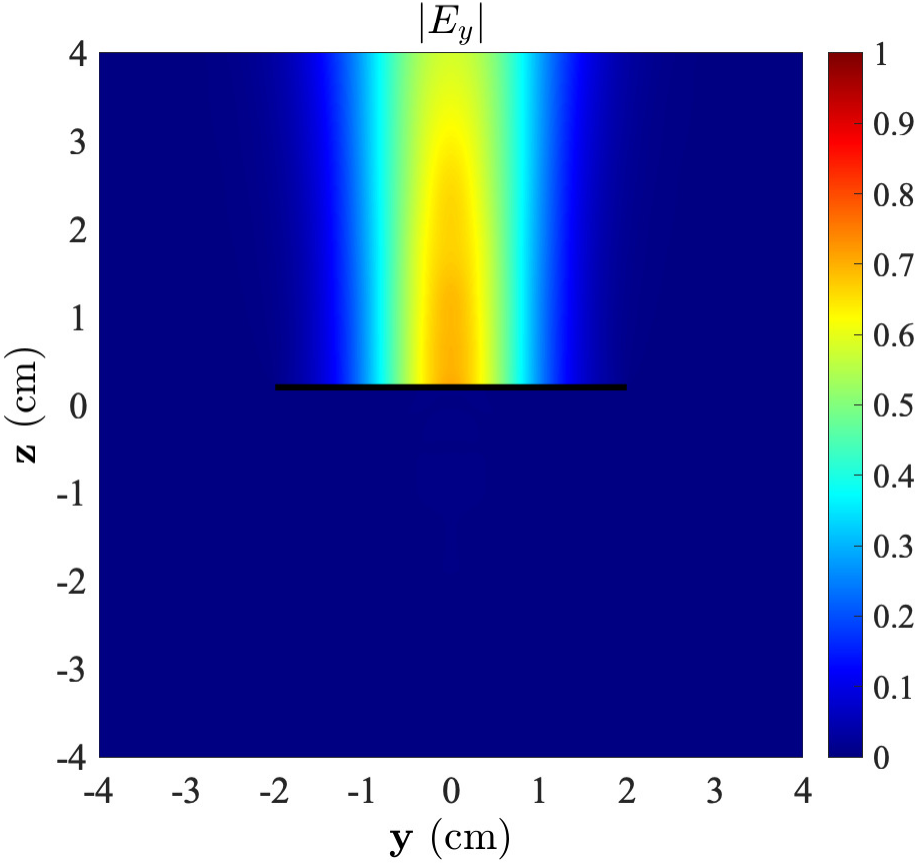}}\vspace{12pt}\\
\subfloat[]{\includegraphics[width=0.6\linewidth]{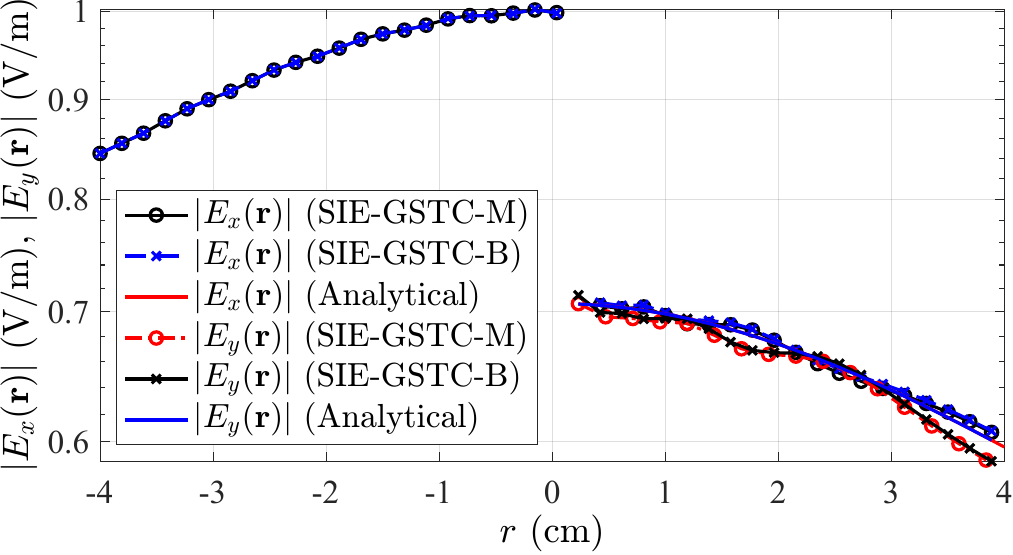}}
\caption{Polarization rotator with $\theta=45^{\circ}$. (a) $|E_x(\mathbf{r})|$ and (b) $|E_y(\mathbf{r})|$ computed using the SIE-GSTC solver on the $yz$-plane for the monoanisotropic realization. (c) $|E_x(\mathbf{r})|$ and $|E_y(\mathbf{r})|$ computed using the SIE-GSTC solver along the $z$-axis, $z \in [-4, 4]\, \mathrm{cm}$, for both the monoanisotropic (``M'') and bianisotropic (``B'') realizations, compared against the analytical expressions.}
\label{Fig:Pol_Rot45}
\end{figure}

\begin{figure}[!ht]
\centering
\subfloat[]{\includegraphics[width=0.46\linewidth]{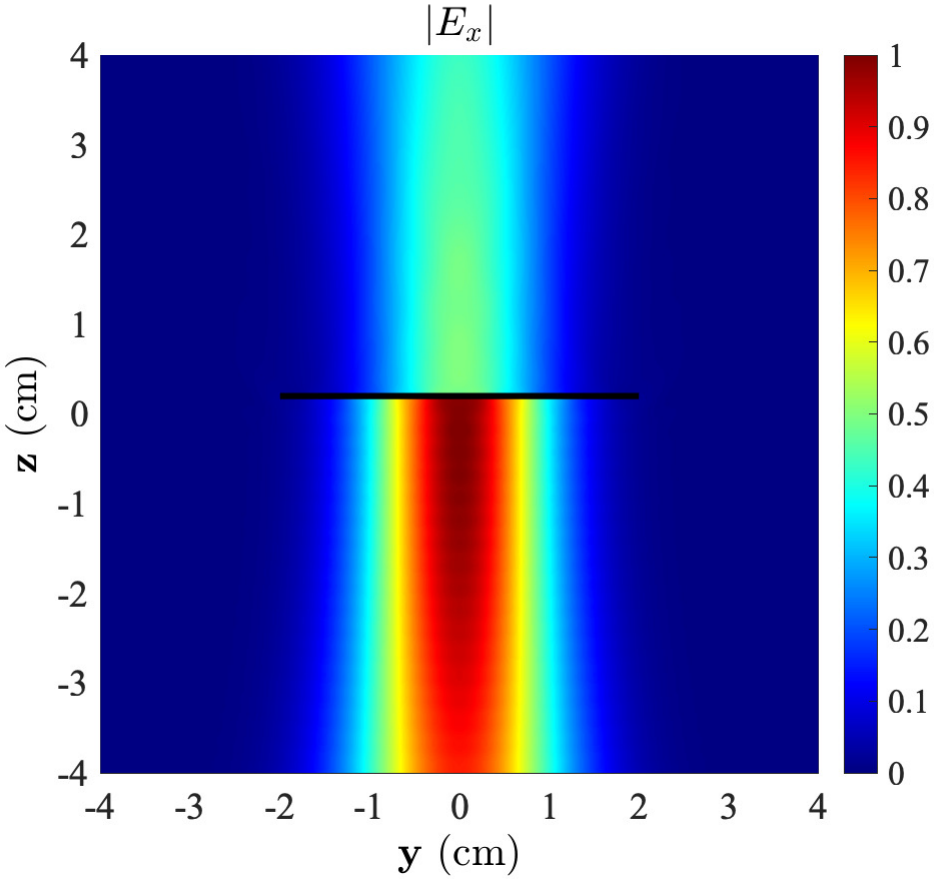}}\hspace{12pt}
\subfloat[]{\includegraphics[width=0.46\linewidth]{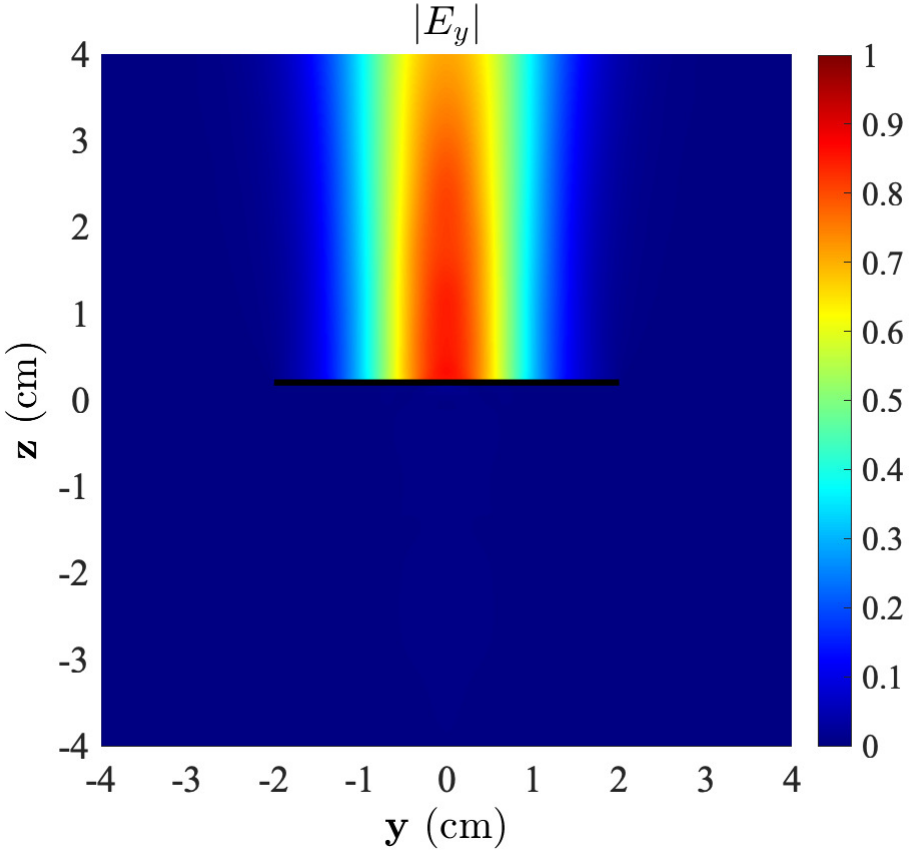}}\vspace{12pt}\\
\subfloat[]{\includegraphics[width=0.6\linewidth]{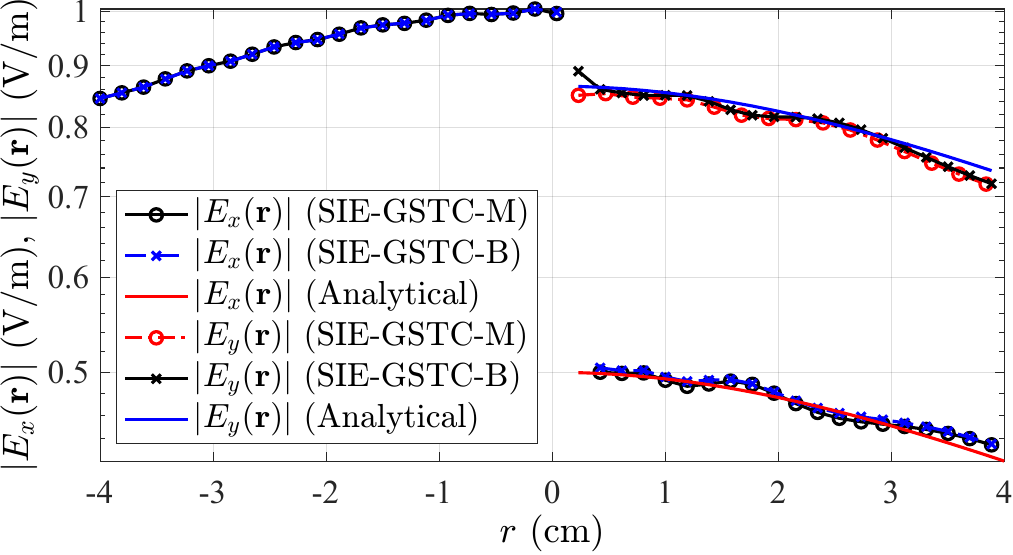}}
\caption{Polarization rotator with $\theta=60^{\circ}$. (a) $|E_x(\mathbf{r})|$ and (b) $|E_y(\mathbf{r})|$ computed using the SIE-GSTC solver on the $yz$-plane for the monoanisotropic realization. (c) $|E_x(\mathbf{r})|$ and $|E_y(\mathbf{r})|$ computed using the SIE-GSTC solver along the $z$-axis, $z \in [-4, 4]\, \mathrm{cm}$, for both the monoanisotropic (``M'') and bianisotropic (``B'') realizations, compared against the analytical expressions.}
\label{Fig:Pol_Rot60}
\end{figure}

Figs.~\ref{Fig:Pol_Rot30},~\ref{Fig:Pol_Rot45}, and~\ref{Fig:Pol_Rot60} present the results for $\theta = 30^{\circ}$, $45^{\circ}$, and $60^{\circ}$, respectively. In each figure, (a)--(b) show the magnitude of the electric field's $x$- and $y$-components, $|E_x(\mathbf{r})|$ and $|E_y(\mathbf{r})|$, computed using the SIE-GSTC solver on the $yz$-plane for the monoanisotropic realization. The figures illustrate the rotation of the incident Gaussian beam upon transmission through the metasurface. The corresponding field plots for the bianisotropic realization are omitted to avoid redundancy. In each figure, (c) compares $|E_x(\mathbf{r})|$ and $|E_y(\mathbf{r})|$ computed using the SIE-GSTC solver along the $z$-axis, $z\in [-4, 4]\,\mathrm{cm}$, for the monoanisotropic (``M'') and bianisotropic (``B'') realizations to those obtained from the analytical expressions derived by applying the prescribed wave transformation~\eqref{eq:transform_rot} to $\mathbf{E}^{\mathrm{inc}}(\mathbf{r})$. Excellent agreement is observed across both realizations.

To quantify this agreement, $\mathrm{err}_{\ell_2}$ is computed using~\eqref{eq:error_metric} along the axial observation line $\mathbf{r}_p = \hat{\mathbf{z}}[z_0 + (p-1)\Delta z]$ with $z_0 = 1.7\,\mathrm{cm}$, $\Delta z = 0.0063\,\mathrm{cm}$, and $N_{\mathrm{p}} = 2101$. For $\theta \in \{30^{\circ}, 45^{\circ}, 60^{\circ}\}$, $\mathrm{err}_{\ell_2} \in \{8.88, 13.07, 16.99\} \times 10^{-3}$ for the monoanisotropic realization and $\mathrm{err}_{\ell_2} \in \{8.89, 12.72, 16.71\} \times 10^{-3}$ for the bianisotropic realization. The error levels are comparable for both realizations and grow modestly with $\theta$.

The convergence behavior of the matrix systems for the two realizations differs significantly. Across the three rotation angles, TFQMR converges in $N_{\mathrm{it}} \in \{2, 2, 2\}$ iterations for the monoanisotropic realization and $N_{\mathrm{it}} \in \{10, 440, 582\}$ iterations for the bianisotropic realization, despite the nonzero susceptibility components having identical magnitudes in the two realizations. The disparity originates in the block structure of $\bar{\bar{Z}}$, specifically in where the susceptibility coupling places the $\mathcal{L}[\mathbf{X}](\mathbf{r})$ contributions relative to the well-tested identity (Gram) term on the diagonal~\cite{Pasi_TN}. For the planar metasurface with purely tangential susceptibilities, the $\tilde{\mathcal{K}}[\mathbf{X}](\mathbf{r})$ contributions vanish under Galerkin testing. For the monoanisotropic realization ($\bar{\bar{\chi}}_{\mathrm{em}}(\mathbf{r})=\bar{\bar{\chi}}_{\mathrm{me}}(\mathbf{r})=\bar{\bar{0}}$), the system reduces to a block-diagonal form in which the diagonal blocks $\bar{\bar{Z}}_{\mathrm{JJ}}$ and $\bar{\bar{Z}}_{\mathrm{MM}}$ carry the well-tested identity (Gram) term together with the $\mathcal{L}[\mathbf{X}](\mathbf{r})$ contribution weighted by $\bar{\bar{\chi}}_{\mathrm{ee}}(\mathbf{r})$ and $\bar{\bar{\chi}}_{\mathrm{mm}}(\mathbf{r})$, respectively. The identity term dominates each block, keeping the system well-conditioned~\cite{Pasi_TN}. For the bianisotropic realization ($\bar{\bar{\chi}}_{\mathrm{ee}}(\mathbf{r})=\bar{\bar{\chi}}_{\mathrm{mm}}(\mathbf{r})=\bar{\bar{0}}$), the $\mathcal{L}[\mathbf{X}](\mathbf{r})$ contribution appears only in the off-diagonal blocks $\bar{\bar{Z}}_{\mathrm{JM}}$ and $\bar{\bar{Z}}_{\mathrm{MJ}}$, weighted by $\bar{\bar{\chi}}_{\mathrm{em}}(\mathbf{r})$ and $\bar{\bar{\chi}}_{\mathrm{me}}(\mathbf{r})$, respectively, coupling $\bar{J}$ and $\bar{M}$. These off-diagonal blocks carry no identity term to control their spectrum, and the resulting spectral spread of $\bar{\bar{Z}}$ degrades TFQMR convergence.

\subsubsection{Curved Metasurface}\label{sec:pol_rot_curved}
For the curved metasurface, the mesh of the planar disk of radius $R_{\mathrm{d}}=4 \lambda_0$ is deformed so that its nodes lie on a spherical cap of curvature radius $R_{\mathrm{c}}=5\lambda_0$, yielding the mesh of the curved surface $S$. The deformation keeps the rim in its original plane and bulges the surface upward toward the apex: each mesh node at radial position $r=\sqrt{x^{2}+y^{2}}$ is displaced along $\hat{\mathbf{z}}$ by
\begin{equation}
  h(r) = \sqrt{R_{\mathrm{c}}^{2}-r^{2}} - \sqrt{R_{\mathrm{c}}^{2}-R_{\mathrm{d}}^{2}}
\end{equation}
which vanishes at the rim ($r=R_{\mathrm{d}}$) and reaches its maximum at the apex ($r=0$), $h_{\max} = R_{\mathrm{c}} - \sqrt{R_{\mathrm{c}}^{2}-R_{\mathrm{d}}^{2}} = 2\lambda_0$.

The same susceptibility tensors synthesized for the planar metasurface at $\theta=30^{\circ}$, for both the monoanisotropic and bianisotropic realizations, are imposed on the curved metasurface. On each triangular patch of the mesh, they are reproduced in the local tangent frame, which is defined relative to the patch normal and the in-plane reference direction obtained by projecting the incident polarization $\hat{\mathbf{x}}$ onto the patch, and then transformed to global Cartesian coordinates. In this way, each patch reproduces the planar design in its own tangent frame, while the global tensor components vary with the patch orientation. 

The curved $S$ is illuminated by the Gaussian beam in~\eqref{eq:GaussInc} with polarization $\hat{\mathbf{p}}=\hat{\mathbf{x}}$ and the beam radius $w_{\mathrm{b}}=2\lambda_0$ at an excitation frequency of $f = 60\,\mathrm{GHz}$. The resulting matrix systems for the monoanisotropic and bianisotropic realizations are solved using TFQMR. 

\begin{figure}[!ht]
\centering
\subfloat[]{\includegraphics[width=0.46\linewidth]{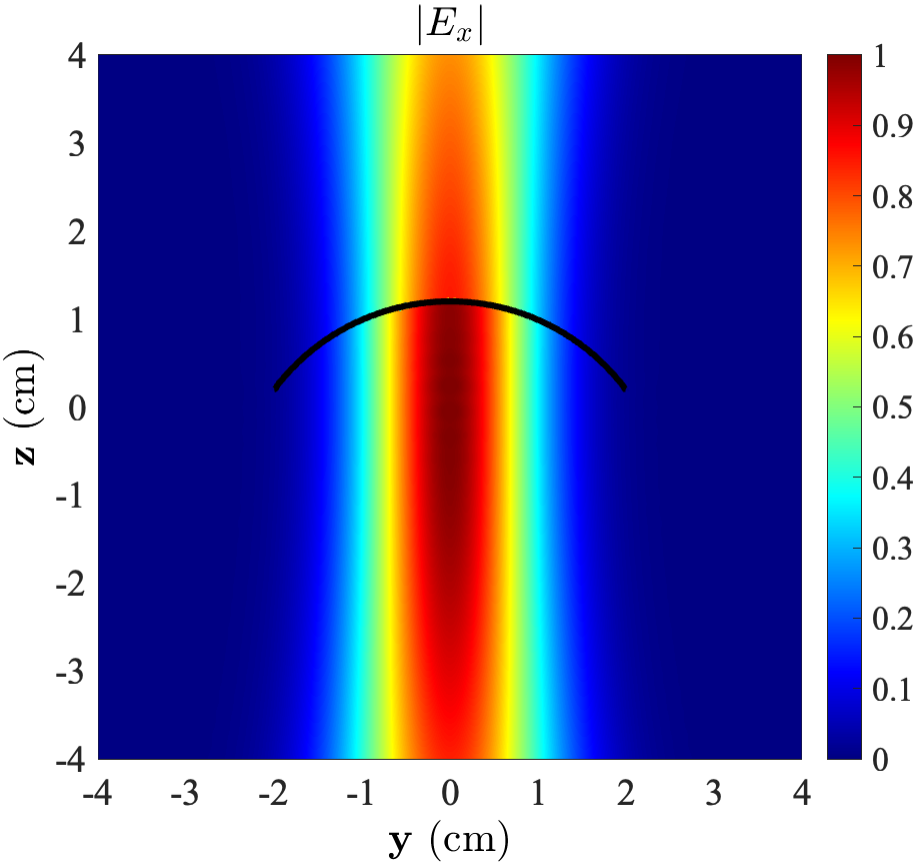}}\hspace{12pt}
\subfloat[]{\includegraphics[width=0.46\linewidth]{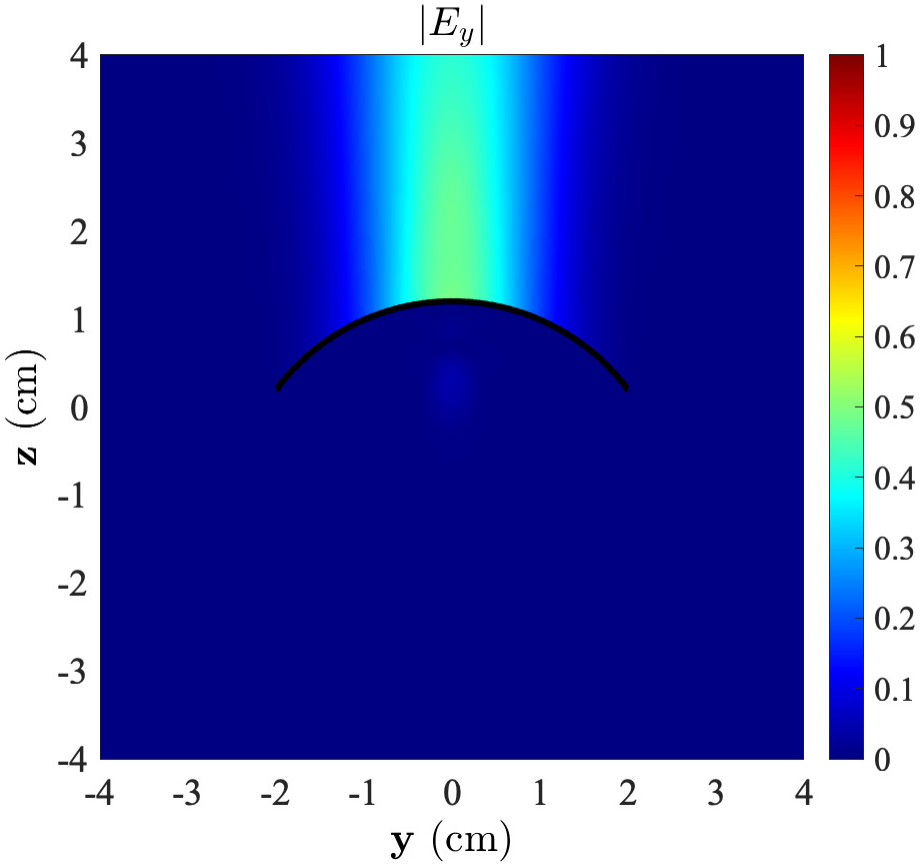}}
\caption{Curved polarization rotator with $\theta=30^{\circ}$. (a) $|E_x(\mathbf{r})|$ and (b) $|E_y(\mathbf{r})|$ computed using the SIE-GSTC solver on the $yz$-plane for the monoanisotropic realization.}
\label{Fig:Rot_Curved_Fields}
\end{figure}

\begin{figure}[!ht]
\centering
\subfloat[]{\includegraphics[width=0.46\linewidth]{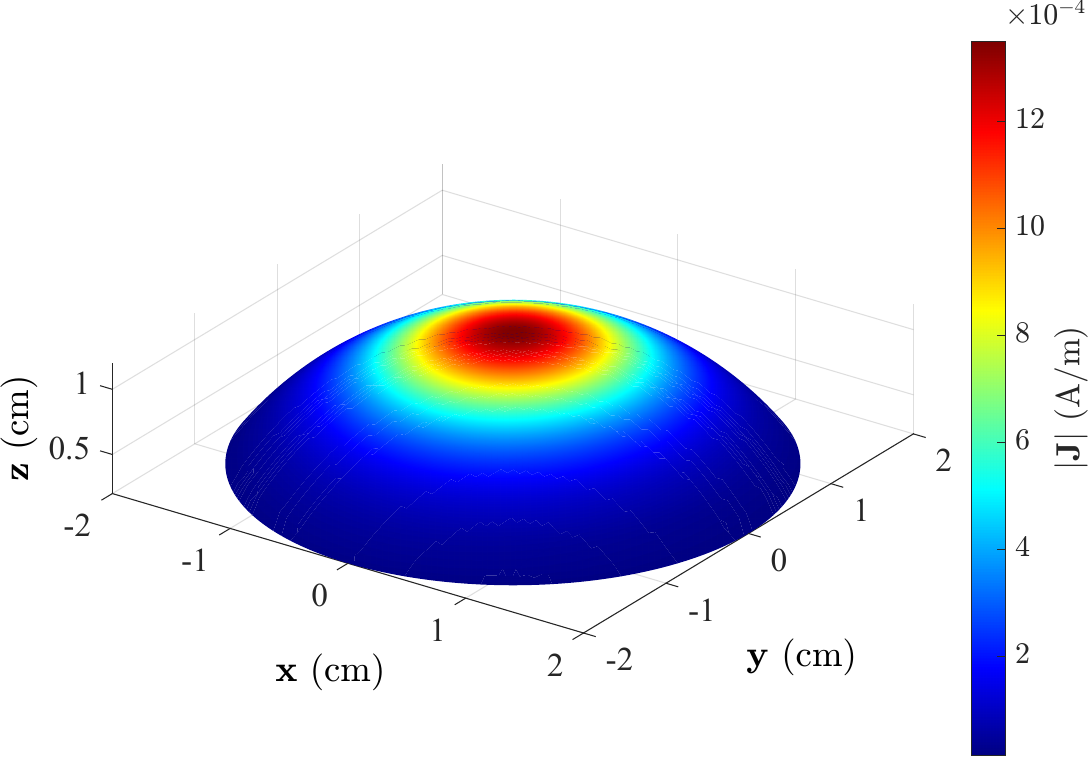}}\hspace{12pt}
\subfloat[]{\includegraphics[width=0.46\linewidth]{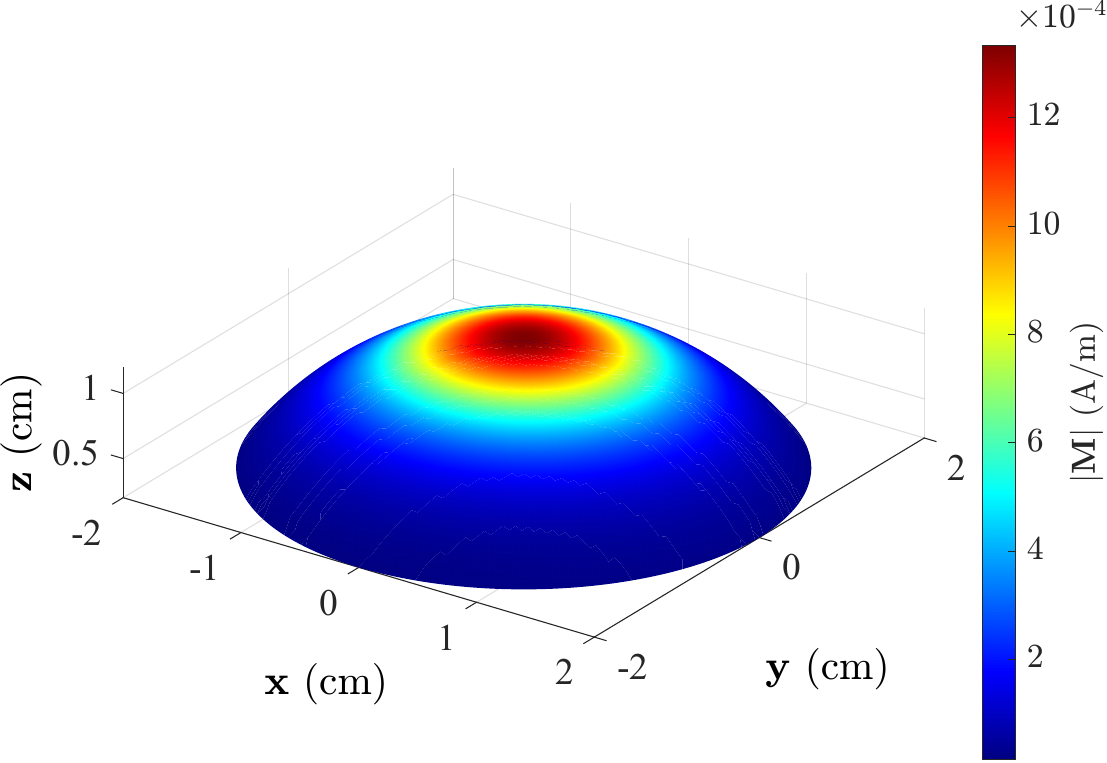}}
\caption{Curved polarization rotator with $\theta=30^{\circ}$. (a) $|\mathbf{J}_{\Sigma}(\mathbf{r})|$ and (b) $|\mathbf{M}_{\Sigma}(\mathbf{r})|$ on the curved surface $S$ for the monoanisotropic realization.}
\label{Fig:Rot_Curved_Currents}
\end{figure}

Fig.~\ref{Fig:Rot_Curved_Fields} shows $|E_x(\mathbf{r})|$ and $|E_y(\mathbf{r})|$ computed using the SIE-GSTC solver on the $yz$-plane for the monoanisotropic realization. Fig.~\ref{Fig:Rot_Curved_Currents} shows the magnitude of the equivalent currents, $|\mathbf{J}_{\Sigma}(\mathbf{r})|$ and $|\mathbf{M}_{\Sigma}(\mathbf{r})|$, on the curved surface $S$. 

The relative error $\mathrm{err}_{\ell_2}$ is computed using~\eqref{eq:error_metric} along the axial observation line $\mathbf{r}_p = \hat{\mathbf{z}}[z_0+(p-1)\Delta z]$ with $z_0=1.99\,\mathrm{cm}$, $\Delta z=0.002\,\mathrm{cm}$, and $N_{\mathrm{p}}=2101$, where the analytical solution obtained for the planar metasurface is used as the reference field $\mathbf{E}_{\mathrm{ref}}(\mathbf{r})$. Its values are $\mathrm{err}_{\ell_2}=8.01\times 10^{-3}$ for the monoanisotropic realization and $\mathrm{err}_{\ell_2}=1.26\times10^{-2}$ for the bianisotropic realization.

TFQMR converges in $N_{\mathrm{it}}=3$ iterations for the monoanisotropic realization and $N_{\mathrm{it}}=34$ for the bianisotropic realization. The monoanisotropic count is comparable to that of the planar metasurface at the same angle, confirming that convergence is preserved once curvature renders the $\tilde{\mathcal{K}}[\mathbf{X}](\mathbf{r})$ contributions nonzero.

\subsection{Perfect Reflection}\label{sec:perfect_reflection}
The metasurface considered in this section is designed to totally reflect a normally incident plane wave, so that no field is transmitted and the reflected field is phase-reversed relative to the incident field. The prescribed field transformation specifies fields on the incident side, $\mathbf{E}(\mathbf{r}_{-})$, and the transmitted side, $\mathbf{E}(\mathbf{r}_{+})$, as
\begin{equation}
\label{eq:transform_reflect}
\begin{aligned}
\mathbf{E}(\mathbf{r}_{-}) & = \hat{\mathbf{x}} (e^{-\mathrm{j}k_0 z} - e^{\mathrm{j}k_0 z})\\
\mathbf{E}(\mathbf{r}_{+}) & = \mathbf{0}.
\end{aligned}
\end{equation}
For this transformation, only a bianisotropic realization is considered; the susceptibility tensors are synthesized assuming full bianisotropy and satisfying the reciprocity condition $\bar{\bar{\chi}}_{\mathrm{me}}(\mathbf{r}) = -\bar{\bar{\chi}}_{\mathrm{em}}^{\mathsf{T}}(\mathbf{r})$~\cite{Lorentz_Reciprocity}. The nonzero components of the resulting tensors are
\begin{equation}
\begin{aligned}
    \chi_{\mathrm{ee}}^{xx}(\mathbf{r})&={4 \mathrm{j}(e^{2 \mathrm{j} k_0 z_{\mathrm{s}}}+1)}/[{k_0(e^{2 \mathrm{j} k_0 z_{\mathrm{s}}}-1)}]\\
     \chi_{\mathrm{em}}^{xy}(\mathbf{r})&={2\mathrm{j}}/{k_0}\\
      \chi_{\mathrm{me}}^{yx}(\mathbf{r})&=-{2\mathrm{j}}/{k_0}
\end{aligned}    
\end{equation}
where $z_{\mathrm{s}}=0.2\,\mathrm{cm}$ is the $z$-coordinate of the metasurface. 

The SIE-GSTC solver is validated by applying the synthesized susceptibility tensors to the same planar metasurface used in Section~\ref{sec:pol_rot_flat}. Surface $S$ is discretized using the same triangular mesh as in Section~\ref{sec:pol_rot_flat}. It is illuminated by the Gaussian beam in~\eqref{eq:GaussInc} with polarization $\hat{\mathbf{p}}=\hat{\mathbf{x}}$ and the beam radius $w_{\mathrm{b}}=2\lambda_0$ at an excitation frequency of $f = 60\,\mathrm{GHz}$. The resulting matrix system is solved using GMRES. 

\begin{figure}[!ht]
\centering
\subfloat[]{\includegraphics[width=0.46\linewidth]{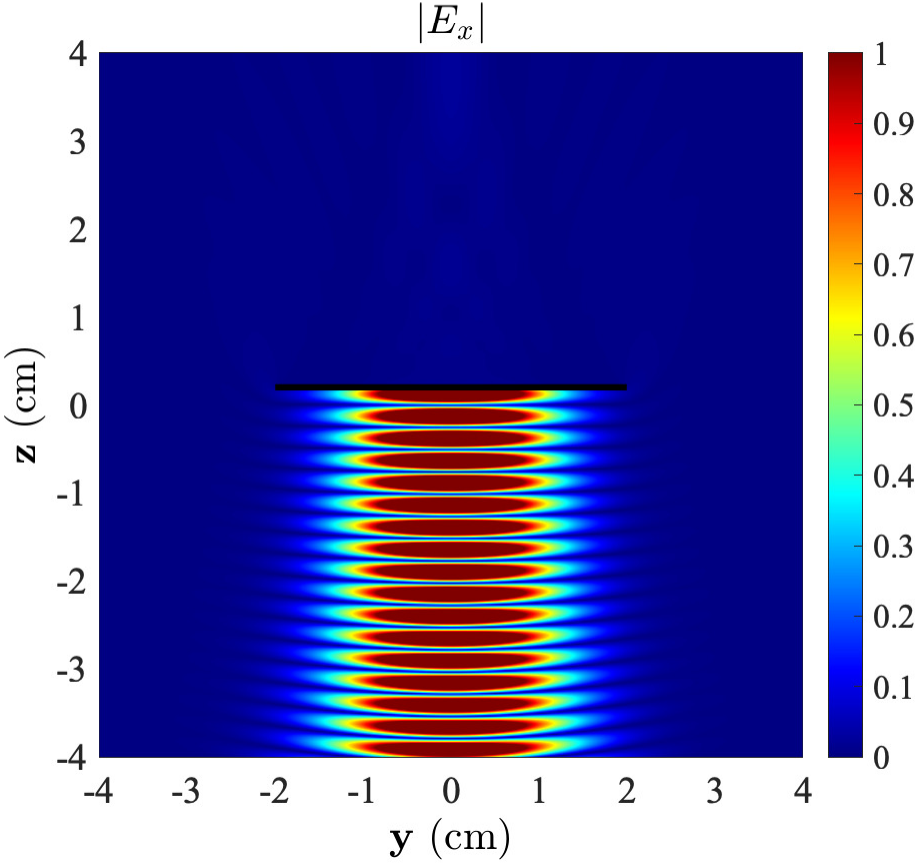}}\\
\subfloat[]{\includegraphics[width=0.6\linewidth]{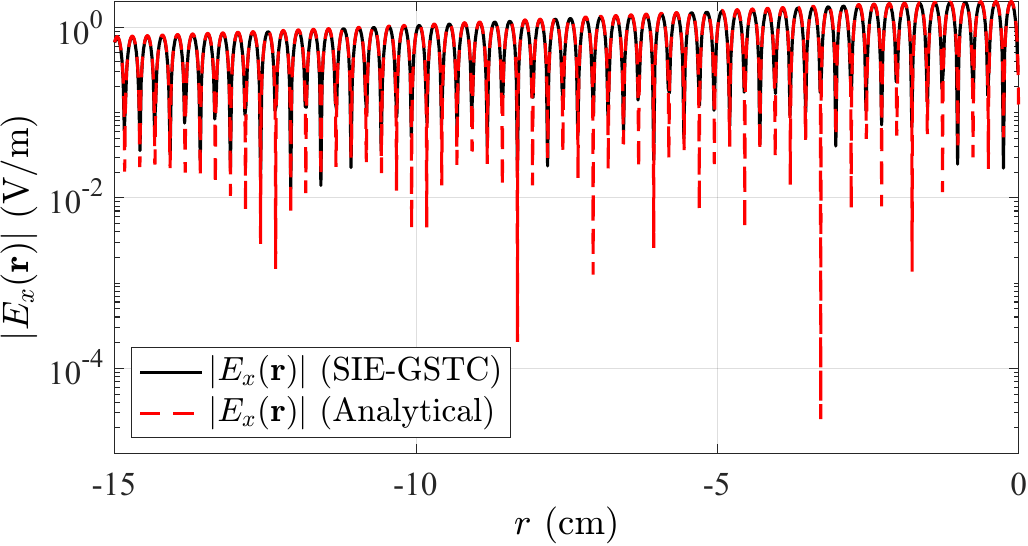}}
\caption{Perfect reflector. (a) $|E_x(\mathbf{r})|$ computed using the SIE-GSTC solver on the $yz$-plane. (b) $|E_x(\mathbf{r})|$ computed using the SIE-GSTC solver along the $z$-axis, $z \in [-15, 0]\,\mathrm{cm}$, compared against the analytical expression.}
\label{Fig:PEC}
\end{figure}

Fig.~\ref{Fig:PEC}(a) shows $|E_x(\mathbf{r})|$ computed using the SIE-GSTC solver on the $yz$-plane, confirming the expected strong reflection of the incident beam. Fig.~\ref{Fig:PEC}(b) compares $|E_x(\mathbf{r})|$ computed using the SIE-GSTC solver along the $z$-axis, $z\in[-15,0]\,\mathrm{cm}$, to that obtained from the analytical expression derived by applying the prescribed wave transformation~\eqref{eq:transform_reflect} to $\mathbf{E}^{\mathrm{inc}}(\mathbf{r})$. Good agreement is observed.

The relative error $\mathrm{err}_{\ell_2}$ is computed using~\eqref{eq:error_metric} along the axial observation line $\mathbf{r}_p = \hat{\mathbf{z}}[z_0 + (p-1)\Delta z]$ with $z_0 = -15\,\mathrm{cm}$, $\Delta z = 0.005\,\mathrm{cm}$, and $N_{\mathrm{p}} = 2990$, yielding $\mathrm{err}_{\ell_2} = 3.96\times10^{-2}$. 

For this transformation, TFQMR stagnated after $1\,000$ iterations with a relative residual of $2.65\times10^{-3}$, failing to reach the prescribed tolerance. This is attributed to the larger susceptibility values required for perfect reflection, which produce stronger field discontinuities and poorer matrix conditioning than the transmissive transformations in Section~\ref{sec:pol_rot}. GMRES was adopted for its greater robustness on poorly conditioned nonsymmetric systems~\cite{GMRES} and converges in $N_{\mathrm{it}}=7$ iterations.

\subsection{Broadband Absorber}\label{sec:broadband}
\begin{figure}[!ht]
\centering
\includegraphics[width=0.6\linewidth]{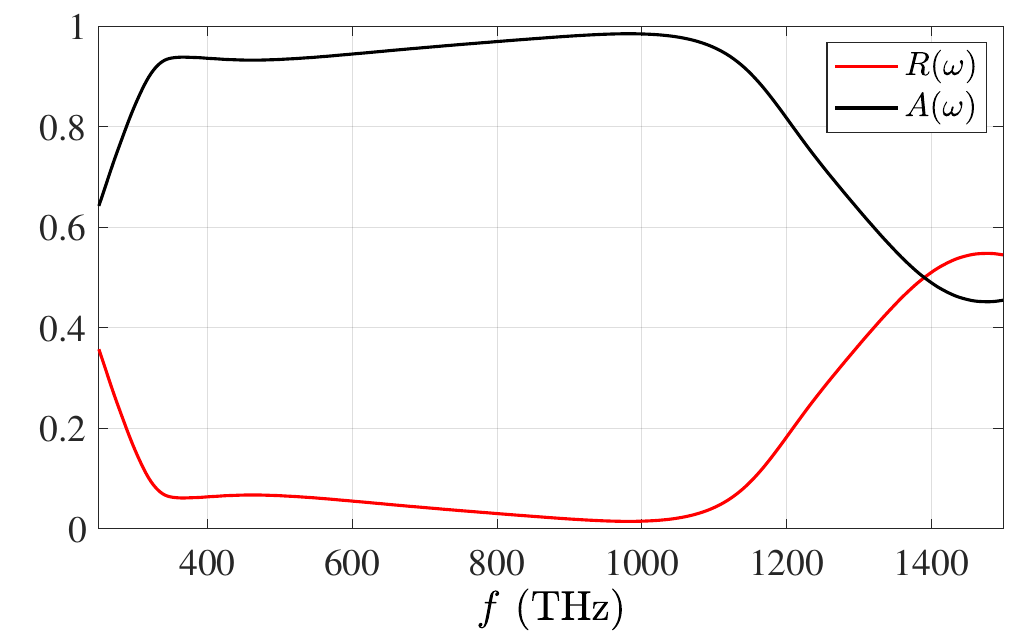}
\caption{Reflectance and absorptance coefficients, $R(\omega)$ and $A(\omega)$, of the broadband absorber derived from scattered fields computed using FDTD~\cite{Mondal2025_PlasmonicAbsorber}.}
\label{fig:R_and_A}
\end{figure}

Broadband absorbers are optical devices engineered to absorb light over a wide spectral range. Such metasurfaces find extensive applications in solar energy harvesting, thermal emitters, imaging systems, and advanced photovoltaic devices~\cite{wei2025metasurface, shin2018thermoplasmonic}. The device described in~\cite{Mondal2025_PlasmonicAbsorber} is a thin-film silicon solar cell designed to be insensitive to the polarization of the incident field. The frequency-dependent field reflection coefficient, $\Gamma(\omega)$, is derived from the scattered fields computed using a full-wave FDTD solver. The corresponding reflectance and absorptance are $R(\omega) = |\Gamma(\omega)|^{2}$ and $A(\omega) = 1 - R(\omega)$, respectively (Fig.~\ref{fig:R_and_A}).

The metasurface considered in this section is designed to reproduce the broadband absorption response of this device, retrieved from the scattered fields computed using a full-wave FDTD solver rather than a closed-form transformation. Based on this response, the susceptibility tensors are synthesized for an opaque boundary condition, prescribed on the incident side, $\mathbf{E}(\mathbf{r}_{-})$, and the transmitted side,  $\mathbf{E}(\mathbf{r}_{+})$, as
\begin{equation}
\label{eq:transform_absorber}
\begin{aligned}
    \mathbf{E}(\mathbf{r}_{-}) &= \hat{\mathbf{p}}(e^{-\mathrm{j}k_{0}z} - \Gamma(\omega) e^{\mathrm{j}k_{0}z}) \\
    \mathbf{E}(\mathbf{r}_{+}) &= \mathbf{0}
\end{aligned}    
\end{equation}
where $\hat{\mathbf{p}} \in \{\hat{\mathbf{x}},\hat{\mathbf{y}}\}$. By solving the prescribed transformation for both orthogonal incident field polarizations,  the nonzero susceptibility components are
\begin{equation*}
\begin{aligned}
\chi^{xx}_{\mathrm{ee}}(\mathbf{r}) &= 2\mathrm{j}(\Gamma(\omega) e^{2\mathrm{j}k_0 z_{\mathrm{s}}} + 1)/(k_0[\Gamma(\omega) e^{2\mathrm{j}k_0 z_{\mathrm{s}}}-1]) \\
\chi^{yy}_{\mathrm{ee}}(\mathbf{r}) &=2\mathrm{j}(\Gamma(\omega) e^{2\mathrm{j}k_0 z_{\mathrm{s}}}+ 1)/(k_0[\Gamma(\omega) e^{2\mathrm{j}k_0 z_{\mathrm{s}}}-1])\\
\chi^{xy}_{\mathrm{me}}(\mathbf{r}) &= 2\mathrm{j}/k_0 \\
\chi^{yx}_{\mathrm{me}}(\mathbf{r}) &= -2\mathrm{j}/k_0
\end{aligned}
\end{equation*}
where $z_{\mathrm{s}}$ is the $z$-coordinate of the metasurface. Unlike the bianisotropic realizations in the previous examples, the Lorentz reciprocity condition is not enforced in this formulation. Moreover, the chosen susceptibility profile is not unique: as shown in Section~\ref{sec:pol_rot}, multiple equivalent models may represent the same metasurface. 

The SIE-GSTC solver is validated by applying the synthesized susceptibility tensors to both a planar and a curved metasurface, described in the following two subsections.

\subsubsection{Planar Metasurface}\label{sec:broadband_flat}
For the planar metasurface, $S$ is a circular disk in the $xy$-plane, centered at $(0,0,0.4\lambda_0)$, with a radius of $R_{\mathrm{d}}=4\lambda_0$, and is discretized into a triangular mesh with an average edge length of approximately $\lambda_0/10$, yielding a matrix of dimension $35\,100 \times 35\,100$. Simulations are carried out over the frequency range $f = 262\,\mathrm{THz}$ to $f=1499\,\mathrm{THz}$~\cite{Mondal2025_PlasmonicAbsorber}. At each frequency, $S$ is illuminated by the Gaussian beam in~\eqref{eq:GaussInc} with the beam radius $w_{\mathrm{b}}=2\lambda_0$, considering the polarizations $\hat{\mathbf{p}}=\hat{\mathbf{x}}$ and $\hat{\mathbf{p}}=\hat{\mathbf{y}}$ separately. The matrix system at each frequency is solved using TFQMR. 

\begin{figure}[!ht]
\centering
\subfloat[]{\includegraphics[width=0.46\linewidth]{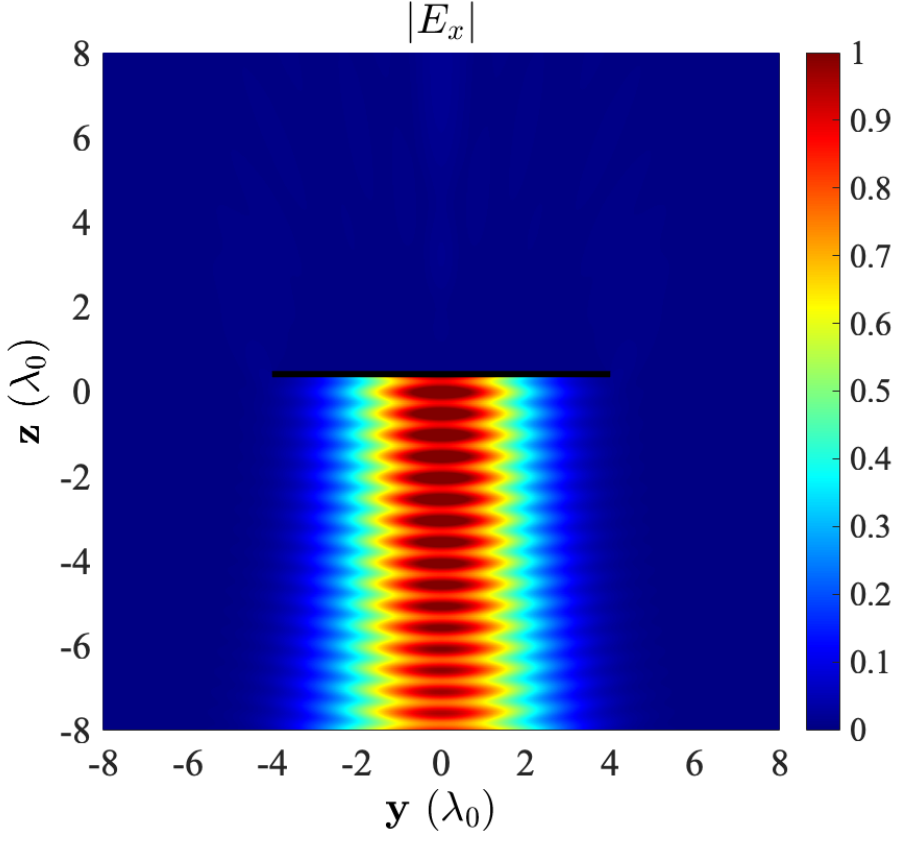}}\hspace{12pt}
\subfloat[]{\includegraphics[width=0.47\linewidth]{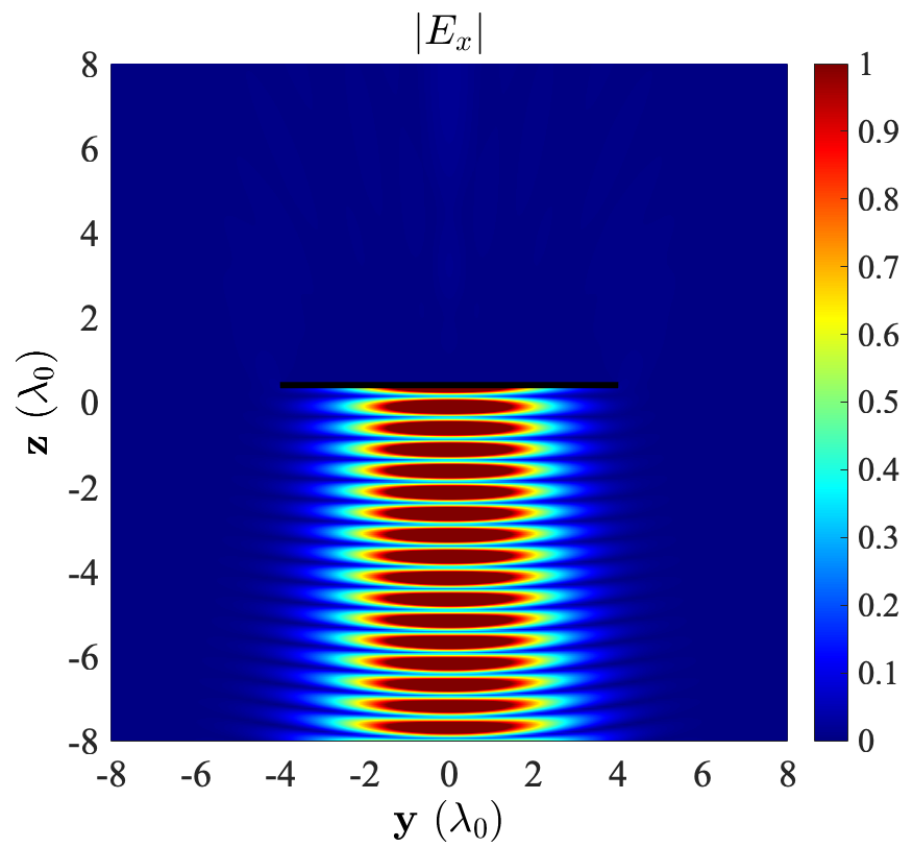}}
\caption{Broadband absorber. $|E_x(\mathbf{r})|$ computed using the SIE-GSTC solver on the $yz$-plane at (a) $f=998\,\mathrm{THz}$ and (b) $f=1499\,\mathrm{THz}$.}
\label{fig:broadband_flat}
\end{figure}

Figs.~\ref{fig:broadband_flat}(a) and~\ref{fig:broadband_flat}(b) show $|E_x(\mathbf{r})|$ computed using the SIE-GSTC solver on the $yz$-plane at $f = 998\,\mathrm{THz}$ and $f = 1499\,\mathrm{THz}$, respectively, for the $\hat{\mathbf{x}}$-polarized incident field. Comparison of these two figures clearly reveals the difference in the field behavior between high-absorption ($f = 998\,\mathrm{THz}$) and high-reflection ($f = 1499\,\mathrm{THz}$) frequencies. 

\begin{figure}[!ht]
\centering
\subfloat[]{\includegraphics[width=0.47\linewidth]{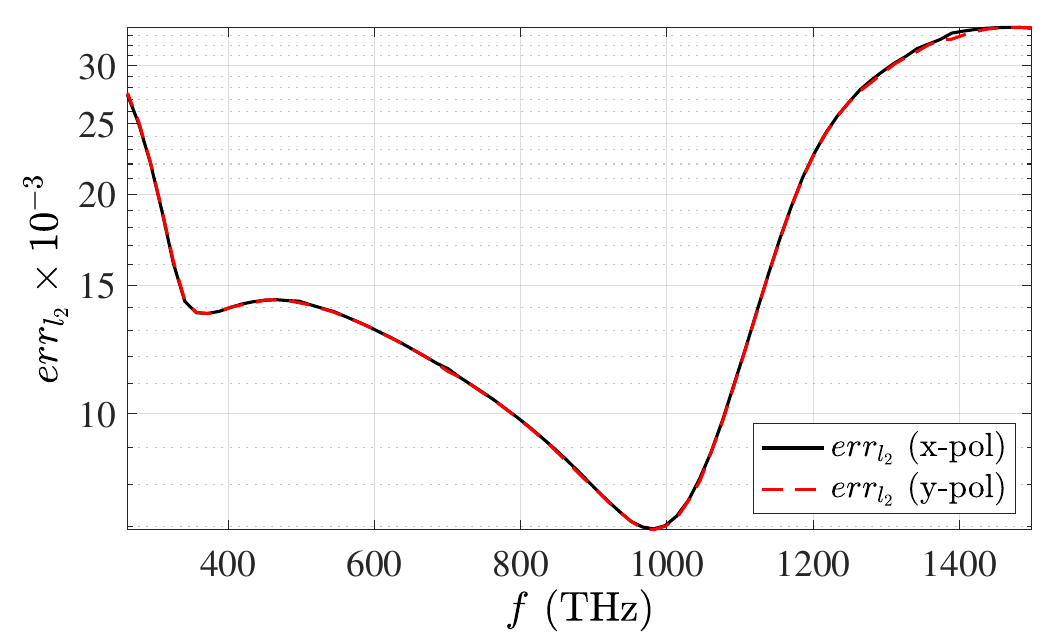}}\hspace{12pt}
\subfloat[]{\includegraphics[width=0.46\linewidth]{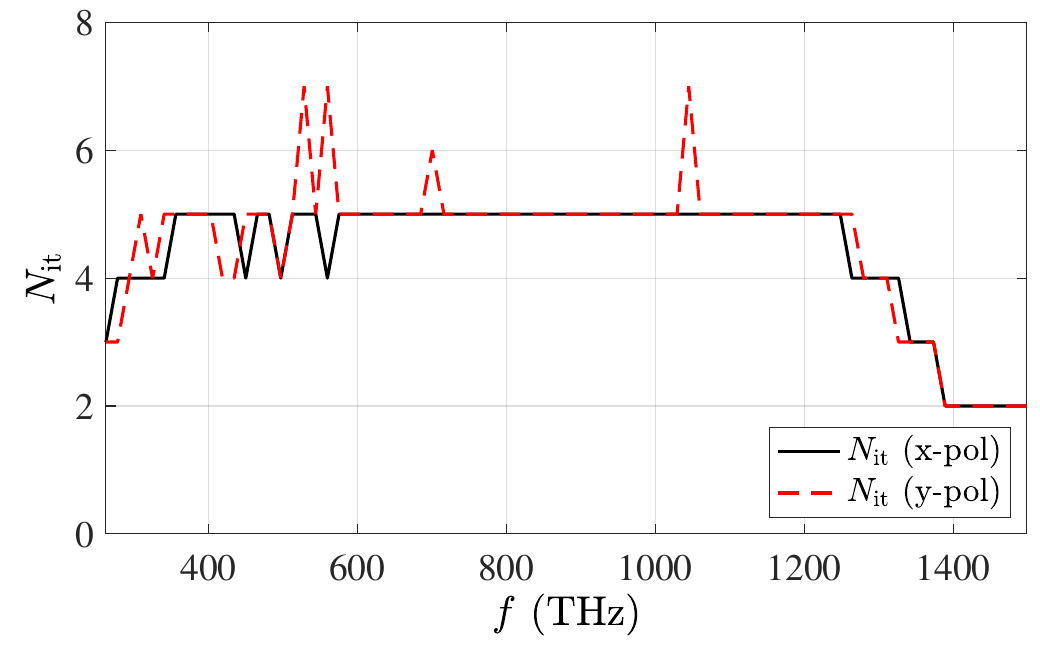}}
\caption{Broadband absorber. (a) Relative error $\mathrm{err}_{\ell_2}$ versus frequency along the axial observation line $\mathbf{r}_p = \hat{\mathbf{z}}[z_0 + (p-1)\Delta z]$, with $z_0 = -16\lambda_0$, $\Delta z = 0.0053\lambda_0$, and $N_{\mathrm{p}} = 2900$, for both incident polarizations. (b) TFQMR iteration count $N_{\mathrm{it}}$ versus frequency for both incident polarizations.}
\label{fig:broadband_error}
\end{figure}

The relative error $\mathrm{err}_{\ell_2}$ is computed using~\eqref{eq:error_metric} along the axial observation line $\mathbf{r}_p = \hat{\mathbf{z}}[z_0 + (p-1)\Delta z]$ with $z_0 = -16\lambda_0$, $\Delta z = 0.0053\lambda_0$, and $N_{\mathrm{p}} = 2900$ in the frequency range $f = 262\,\mathrm{THz}$ to $f=1499\,\mathrm{THz}$ for the $\hat{\mathbf{x}}$- and $\hat{\mathbf{y}}$-polarized incident fields. In the error calculation, $\mathbf{E}_{\mathrm{ref}}(\mathbf{r})$ is analytically obtained by applying the prescribed wave transformation~\eqref{eq:transform_absorber} to $\mathbf{E}^{\mathrm{inc}}(\mathbf{r})$. Fig.~\ref{fig:broadband_error}(a) plots the resulting values of $\mathrm{err}_{\ell_2}$ versus frequency. The results exhibit virtually polarization-insensitive performance. Consistent with the observations in Section~\ref{sec:perfect_reflection}, the error marginally increases for larger values of $R(\omega)$ (higher reflectivity) and decreases otherwise; nevertheless, it remains strictly within acceptable bounds across the entire frequency range.

Fig.~\ref{fig:broadband_error}(b) plots the TFQMR iteration count, $N_{\mathrm{it}}$, versus frequency. $N_{\mathrm{it}}$ stays low for both polarizations.

\subsubsection{Curved Metasurface}\label{sec:broadband_curved}
For the curved metasurface, the mesh of the planar disk is deformed following the same procedure as in Section~\ref{sec:pol_rot_curved}. The broadband susceptibility tensors, synthesized for the planar metasurface, are imposed on this curved mesh in the local tangent frame of each triangular patch, following the same procedure as in Section~\ref{sec:pol_rot_curved}. The curved surface $S$ is excited by the Gaussian beam in~\eqref{eq:GaussInc} with the polarization $\hat{\mathbf{p}}=\hat{\mathbf{x}}$ and the beam radius $w_{\mathrm{b}}=2\lambda_0$ at $f = 998\,\mathrm{THz}$ and $f = 1499\,\mathrm{THz}$. The resulting matrix system at each frequency is solved using TFQMR.

\begin{figure}[!ht]
\centering
\subfloat[]{\includegraphics[width=0.46\linewidth]{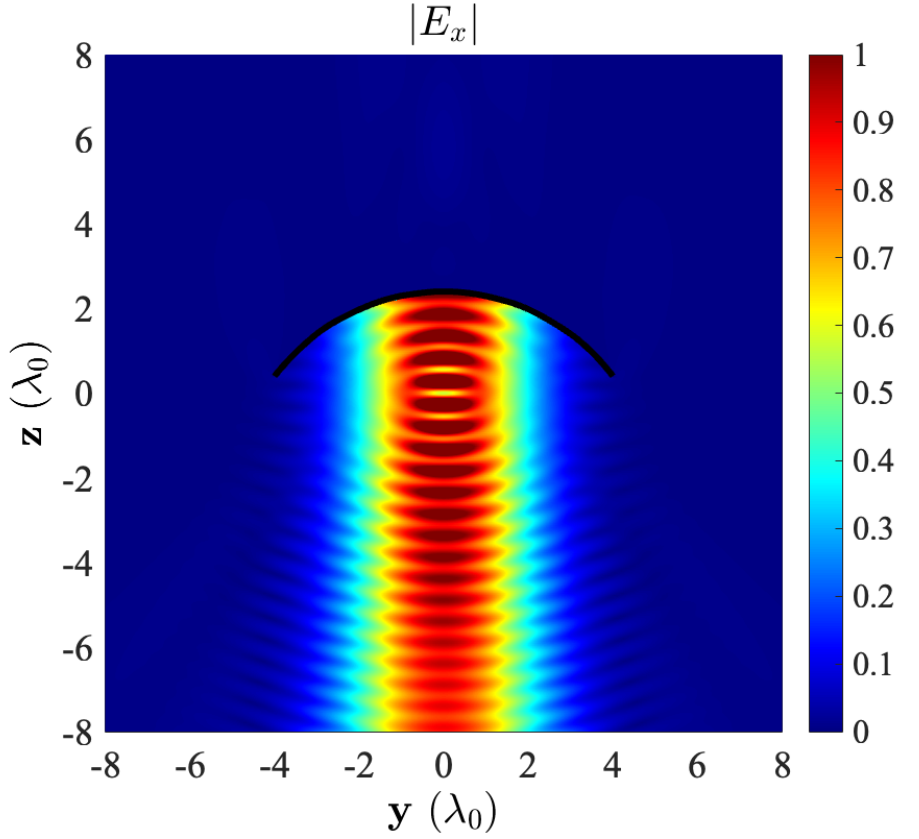}}\hspace{12pt}
\subfloat[]{\includegraphics[width=0.46\linewidth]{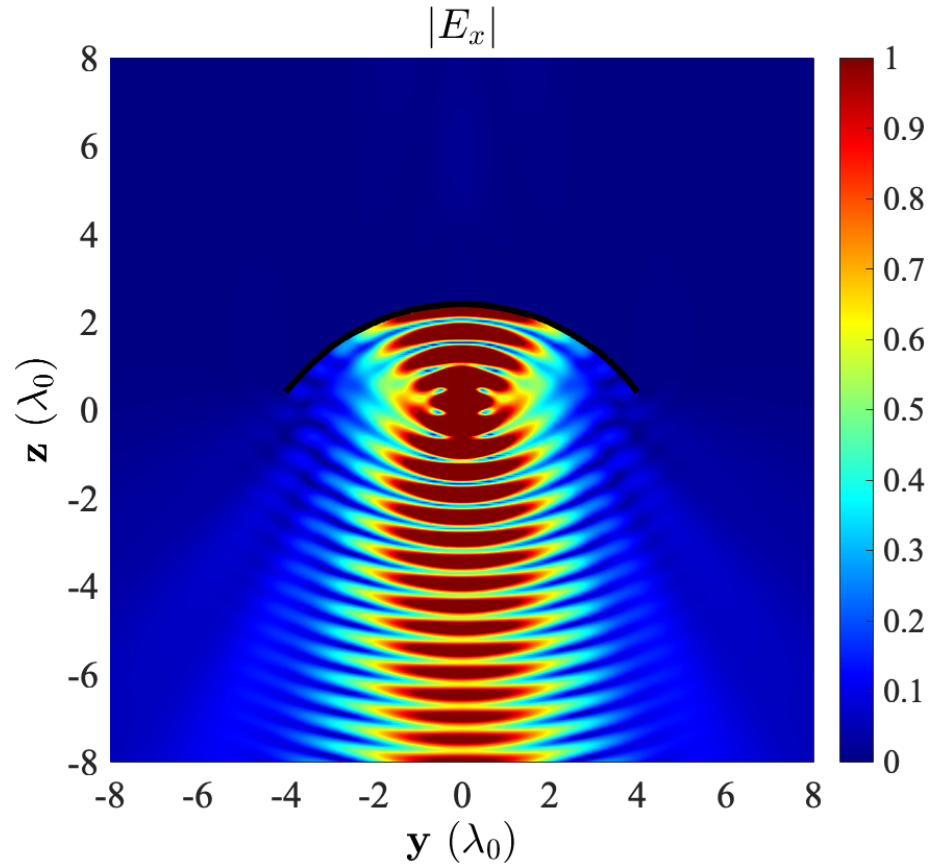}}
\caption{Curved broadband absorber. $|E_x(\mathbf{r})|$ computed using the SIE-GSTC solver on the $yz$-plane at (a) $f=998\,\mathrm{THz}$ and (b) $f=1499\,\mathrm{THz}$.}
\label{fig:broadband_curved}
\end{figure}

Figs.~\ref{fig:broadband_curved}(a) and~\ref{fig:broadband_curved}(b) show $|E_x(\mathbf{r})|$ computed using the SIE-GSTC solver on the $yz$-plane at $f=998\,\mathrm{THz}$ and $f=1499\,\mathrm{THz}$, respectively. The curved-mesh field patterns in Fig.~\ref{fig:broadband_curved} differ visibly from the flat-mesh results in Fig.~\ref{fig:broadband_flat}: because the susceptibilities are synthesized under the normal-incidence assumption, the off-apex regions of the cap intercept the Gaussian beam obliquely and depart from the designed response, producing the additional reflections and interference fringes visible on the illuminated side of the metasurface ($z<0$) in Fig.~\ref{fig:broadband_curved}.

TFQMR converges robustly for the curved metasurface, in $N_{\mathrm{it}}=12$ iterations at $f=998\,\mathrm{THz}$ and $N_{\mathrm{it}}=58$ at $f=1499\,\mathrm{THz}$.

\subsection{Comparison With the Multi-Trace Formulation}\label{sec:multi_trace}
An SIE-GSTC solver that relies on a multi-trace formulation is introduced for the simulation of metasurfaces in~\cite{journal1_Celis_2026}. This formulation defines two independent sets of currents on the interior and exterior sides of a topologically closed surface. These sets are explicitly coupled through the GSTCs. This solver can be used to model an open metasurface by enforcing the nonzero susceptibility tensors only on the active portion of the closed surface while assigning trivial susceptibilities to the remainder. Relative to the present single-trace formulation, the multi-trace construction requires a substantially larger number of unknowns. This subsection quantifies the resulting accuracy and efficiency trade-offs.

To this end, the monoanisotropic realization of an open metasurface designed to rotate the polarization of the incident field by $\theta = 30^{\circ}$ is considered. The expressions of the nonzero susceptibility tensor components are provided in Section~\ref{sec:pol_rot}. For the single-trace formulation, open surface $S$ is a circular disk of radius $2\lambda_0$ centered at $(0,0,0.4\lambda_0)$, and for the multi-trace formulation, closed surface $S$ is a cylinder of radius $2\lambda_0$ and height $0.5\lambda_0$. The top face of the cylinder is centered at $(0,0,0.4\lambda_0)$. The nonzero susceptibility tensors are enforced exclusively on this top face and trivial susceptibility tensors $\bar{\bar{\chi}}_{\mathrm{ee}}(\mathbf{r}) = \bar{\bar{\chi}}_{\mathrm{mm}}(\mathbf{r}) = \bar{\bar{0}}$ are assigned to the remaining faces. In each formulation, $S$ is discretized using a triangular mesh with an average edge length of approximately $\lambda_0/10$, yielding a matrix with dimensions $8\,784 \times 8\,784$ for the single-trace formulation and $44\,748 \times 44\,748$ for the multi-trace formulation. In both formulations, $S$ is illuminated by the Gaussian beam in~\eqref{eq:GaussInc} with polarization $\hat{\mathbf{p}} = \hat{\mathbf{x}}$ and the beam radius $w_{\mathrm{b}} = \lambda_0$ at an excitation frequency of $f = 60\,\mathrm{GHz}$. The resulting matrix systems are solved using TFQMR.

\begin{figure}[!ht]
\centering
\includegraphics[width=0.6\linewidth]{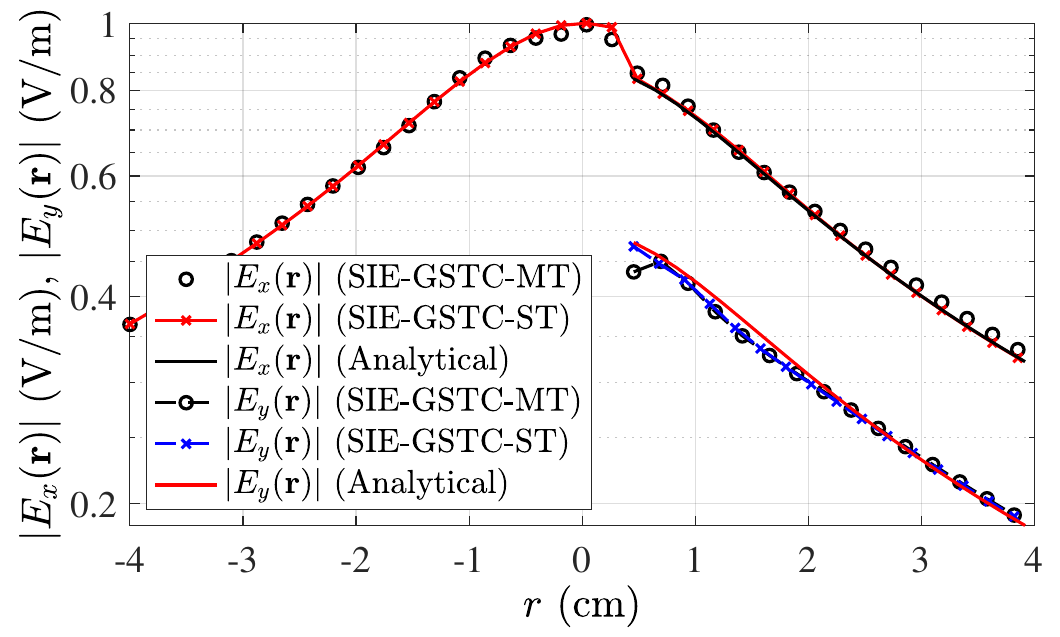}
\caption{Comparison of $|E_x(\mathbf{r})|$ and $|E_y(\mathbf{r})|$ computed using the single-trace (ST) and multi-trace (MT) solvers along the $z$-axis, $z \in [-4, 4]\,\mathrm{cm}$, compared against the analytical expressions.}\label{Fig:N_T_Comparison}
\end{figure}

Fig.~\ref{Fig:N_T_Comparison} compares $|E_x(\mathbf{r})|$ and $|E_y(\mathbf{r})|$ computed using the single-trace solver (ST) and the multi-trace solver (MT) along the $z$-axis for $z \in [-4, 4]\,\mathrm{cm}$, with those obtained from the analytical expressions derived by applying the prescribed wave transformation~\eqref{eq:transform_rot} to the incident field. Very good agreement is observed.

The relative error $\mathrm{err}_{\ell_2}$ is computed using~\eqref{eq:error_metric} along the axial observation line $\mathbf{r}_p = \hat{\mathbf{z}}[z_0 + (p-1)\Delta z]$, with $z_0 = 1.69\,\mathrm{cm}$, $\Delta z = 0.0063\,\mathrm{cm}$, and $N_{\mathrm{p}} = 2101$. The error values are $\mathrm{err}_{\ell_2} = 1.49 \times 10^{-2}$ for the single-trace formulation and 
$\mathrm{err}_{\ell_2} = 2.67 \times 10^{-2}$ for the multi-trace formulation. Both systems remain well-conditioned; TFQMR converges in $N_{\mathrm{it}} = 2$ iterations for the single-trace formulation, compared with $N_{\mathrm{it}} = 8$ for the multi-trace formulation.

For this inherently non-enclosing geometry, the single-trace formulation achieves lower error with roughly five times fewer unknowns than the multi-trace formulation.

\section{Conclusions}\label{sec:Conclusions}
An SIE-GSTC solver is presented for the simulation of 3D open bianisotropic metasurfaces. The metasurface is modeled as an infinitesimally thin, non-enclosing sheet. The proposed approach uses a single set of equivalent surface currents on the sheet, in place of the two sets used by prior multi-trace formulations. The scattered fields on both faces of the sheet, expressed through SIE operators acting on these currents, are substituted into the GSTCs. This system of equations is discretized using RWG basis functions. The resulting solver models an open metasurface directly, without an artificial closure, and applies to both planar and curved geometries. The spatial variation of the susceptibility tensors introduces an additional charge term through their surface divergence, which is incorporated via line integrals along the edges of the mesh triangles.

The solver is validated against analytical solutions for canonical monoanisotropic and bianisotropic transformations, namely polarization rotation and perfect reflection, and is used to model a realistic broadband plasmonic absorber whose susceptibility tensors are retrieved from full-wave simulation data. A direct comparison further shows that the single-trace formulation attains lower error than a multi-trace formulation while using significantly fewer unknowns. 

Future work will focus on the inclusion of normal susceptibility components in the GSTCs, acceleration through the fast multipole method and fast Fourier transform-based algorithms for electrically large deployments, and preconditioning strategies for transformations that yield poorly conditioned systems.

\bibliographystyle{IEEEtran}
\bibliography{references_final.bib}

\newpage\clearpage

\section*{Figures}

\newpage\clearpage

\newpage\clearpage

\newpage\clearpage

\newpage\clearpage

\newpage\clearpage

\newpage\clearpage

\newpage\clearpage


\end{document}